\documentclass{IEEEtran}
\usepackage{epsfig}
\usepackage{amsmath}
\usepackage{amstext}
\usepackage{amsfonts}
\usepackage{amssymb}
\usepackage{eucal}
\usepackage{graphicx}
\usepackage{verbatim}
\usepackage{stmaryrd}
\usepackage{bm}
\usepackage{cite}

\usepackage{tikz}
\usepackage{pgfplots}
\usepackage{algorithmic}

\usetikzlibrary{arrows,petri,topaths}
\usepackage{tkz-berge}

\newcommand{\RR}{{\mathbb{R}}}

\newcommand{\CC}{{\mathbb{C}}}
\newcommand{\herm}{{\sf H}}
\newcommand{\trans}{{\sf T}}

\renewcommand{\P}{{\bf P}}

\newcommand{\R}{{\bf R}}
\newcommand{\X}{{\bf X}}
\newcommand{\Z}{{\bf Z}}
\renewcommand{\S}{{\bf S}}

\newcommand{\Y}{{\bf Y}}
\newcommand{\W}{{\bf W}}

\newcommand{\T}{{\bf T}}

\newcommand{\U}{{\bf U}}

\renewcommand{\H}{{\bf H}}
\newcommand{\I}{{\bf I}}
\newcommand{\E}{{\bf E}}
\newcommand{\x}{{\bf x}}

\newcommand{\h}{{\bf h}}

\renewcommand{\aa}{{\bf a}}
\newcommand{\blambda}{{\bm \lambda}}
\newcommand{\bb}{{\bf b}}
\newcommand{\y}{{\bf y}}

\newcommand{\s}{{\bf s}}
\newcommand{\uu}{{\bf u}}
\newcommand{\e}{{\bf e}}
\newcommand{\w}{{\bf w}}

\newcommand{\oh}{{\frac{1}{2}}}

\newcommand{\uF}{\underline{F}}

\newcommand{\asto}{\overset{\rm a.s.}{\longrightarrow}}

\newcommand{\EE}{{\rm E}}

\DeclareMathOperator{\tr}{tr}
\DeclareMathOperator{\diag}{diag}

\newcounter{ctheorem}
\newtheorem{theorem}[ctheorem]{Theorem}

\newcounter{cdefinition}
\newtheorem{definition}[cdefinition]{Definition}

\begin{document}
\bibliographystyle{IEEEtran}

\title{Signal Processing in Large Systems: \\ A New Paradigm}

\author{Romain~Couillet$^{1}$ and M\'erouane~Debbah$^2$\\ {\it $^1$ Telecommunication Department, Sup\'elec, Gif sur Yvette, France.} \\ {\it $^2$ Alcatel Lucent Chair on Flexible Radio, Gif sur Yvette, Sup\'elec, France.}}

\maketitle

\begin{abstract}
For a long time, detection and parameter estimation methods for signal processing have relied on asymptotic statistics as the number $n$ of observations of a population grows large comparatively to the population size $N$, i.e. $n/N\to \infty$. Modern technological and societal advances now demand the study of sometimes extremely large populations and simultaneously require fast signal processing due to accelerated system dynamics. This results in not-so-large practical ratios $n/N$, sometimes even smaller than one. A disruptive change in classical signal processing methods has therefore been initiated in the past ten years, mostly spurred by the field of large dimensional random matrix theory. The early works in random matrix theory for signal processing applications are however scarce and highly technical. This tutorial provides an accessible methodological introduction to the modern tools of random matrix theory and to the signal processing methods derived from them, with an emphasis on simple illustrative examples. 
\end{abstract}

\section{Introduction}
\label{sec:introduction}
Most signal processing methods, e.g. statistical tests or parameter estimators, are based on asymptotic statistics of $n$ observed random signal vectors \cite{VAN00}. This is because a deterministic behavior often arises when $n\to\infty$, which simplifies the problem as exemplified by the celebrated law of large numbers and the central limit theorem. With the increase of the systems dimension, denoted by $N$, and the need for even faster dynamics, the scenario $n\gg N$ becomes less and less likely to occur in practice. As a consequence, the large $n$ properties are no longer valid, as will be discussed in Section~\ref{sec:signalprocessing}. This is the case for instance in array processing where a large number of antennas is used to detect incoming signals with short time stationarity properties (e.g. radar detection of fast moving objects) \cite{MES08c}. Other examples are found for instance in mathematical finance where numerous stocks show price index correlation over short observable time periods \cite{POT00} or in evolutionary biology where the joint presence of multiple genes in the genotype of a given species is analyzed from a few DNA samples \cite{HAN96}.

This tutorial introduces recent tools to cope with the $n \simeq N$ limitation of classical approaches. These tools are based on the study of large dimensional random matrices, which originates from the works of Wigner \cite{WIG55} in 1955 and, more importantly here, of Mar\u{c}enko and Pastur \cite{MAR67} in 1967. Section~\ref{sec:rmt} presents a brief introduction to random matrix theory. More precisely, we will explain the basic tools used in this field which differ greatly from classical approaches, based on which we will introduce statistical inference tools in the $n \simeq N$ regime. These tools use random matrix theory jointly with complex analysis methods \cite{RUD86} and are often known as G-estimation, named after the G-estimator formulas from Girko \cite{GIR90}. Since the techniques presented in Section~\ref{sec:rmt} are expected to be rather new to the non-expert reader, we will elaborate on toy examples to introduce the methods rather than on a list of theoretical results. More realistic application examples, taken from the still limited literature, are then presented in Section~\ref{sec:applications}. These will further familiarize the reader with the introduced concepts. These examples span from signal detection in array processing to failure localisation in large dimensional networks. Finally, in Section~\ref{sec:conclusion}, we will draw some conclusions and provide future prospects of this emerging field. %In the appendix, we also give a helpful reminder of the mathematical prerequisites needed for better understanding. These prerequisites are mainly on matrix calculus, see e.g. \cite{HOR85,HOR91}, notions of convergence in probability theory, see e.g. \cite{BIL08}, and on basic elements of complex analysis, see e.g. \cite{RUD86}.

{\it Notations:} Boldface lowercase (uppercase) characters stand for vectors (matrices), with $\I_N$ the $N\times N$ identity matrix. The notations $(\cdot)^\trans$ and $(\cdot)^\herm$ denote transpose and Hermitian transpose, respectively. The value $\sqrt{-1}$ is denoted by $\imath$. The notation $\EE$ is the expectation operator. The notations $\asto$ and $\Rightarrow$ denote almost sure and weak convergence of random variables, respectively. The symbol $\mathcal N(\mu,\sigma^2)$ indicates a real Gaussian distribution with mean $\mu$ and variance $\sigma^2$. The norm $\Vert \cdot \Vert$ will be understood as the Euclidean norm for vectors and the spectral norm for Hermitian matrices ($\Vert \X \Vert=\max_i|\lambda_i|$ for $\X$ with eigenvalues $\lambda_1,\ldots,\lambda_N$). The Dirac delta is denoted $\delta(x)$. Finally, $(x)^+=\max(x,0)$ for real $x$, $1_{x\leq y}$ is the function of $x$ equal to zero for $x>y$ and equal to one for $x\leq y$. %, and $\delta_x^y=1_{x\leq y}1_{y\leq x}$. 

\section{Limitations of classical signal processing}
\label{sec:signalprocessing}

We start our exposition with a simple example explaining the compelling need for the study of large dimensional random matrices.

\subsection{The Mar\u{c}enko-Pastur law}
Consider $\y=\T^\oh\x\in\CC^N$ a random variable with $\x$ having complex independent and identically distributed (i.i.d.) entries with zero mean and unit variance. If $\y_1,\ldots,\y_n$ are $n$ independent realizations of $\y$, then, from the strong law of large numbers,
\begin{equation}
	\label{eq:RntoR}
	\left\Vert \frac1n\sum_{i=1}^n \y_i\y_i^\herm - \T\right\Vert \asto 0
\end{equation}
as $n\to\infty$. For further use, we will denote $\Y=[\y_1,\ldots,\y_n]$ with $(i,j)$ entry $Y_{ij}$, and realize that with this notation,
\begin{align*}
	\hat{\T} \triangleq \frac1n\Y\Y^\herm = \frac1n\sum_{i=1}^n \y_i\y_i^\herm.
\end{align*}

Here $\hat{\T}$ is often called the {\it sample covariance matrix}, obtained from the observations $\y_1,\ldots,\y_n$ of the random variable $\y$ with {\it population covariance matrix} $\EE[\y\y^\herm]=\T$. The convergence \eqref{eq:RntoR} suggests that, for $n$ sufficiently large, $\hat{\T}$ is an accurate estimate of $\T$. The natural reason why $\T$ can be asymptotically well approximated is because $\hat{\T}$ originates from a total of $Nn\gg N^2$ observations, while the parameter $\T$ to be estimated is of size $N^2$.

Suppose now for simplicity that $\T=\I_N$. If the relation $Nn\gg N^2$ is not met in practice, e.g. if $n$ cannot be taken very large compared to the system size $N$, then a peculiar behaviour of $\hat{\T}$ arises, as both $N,n\to\infty$ while $N/n$ does not tend to zero. Observe that the entry $(i,j)$ of $\hat{\T}$ is given by
\begin{equation*}
	[\hat{\T}]_{ij} = \frac1n\sum_{k=1}^n Y_{ik}Y^\ast_{jk}.
\end{equation*}

Since $\T=\I_N$, $\EE[Y_{ik}Y^\ast_{jk}]=\delta_{ij}$. Therefore, the law of large numbers ensures that $[\hat{\T}]_{ij}\to 0$ if $i\neq j$, while $[\hat{\T}]_{ii}\to 1$, for all $i,j$. Since $N$ grows along with $n$, we might then say that $\hat{\T}$ converges point-wise to an ``infinitely large identity matrix''. This stands, irrelevant of $N$, as long as $n\to\infty$, so in particular if $N= 2n \to \infty$. However, under this hypothesis, $\hat{\T}$ is of maximum rank $N/2$ and is therefore rank-deficient.\footnote{Indeed, $\hat{\T}=\frac2N\sum_{i=1}^{N/2}\y_i\y_i^\herm$, which is the sum of $N/2$ rank-$1$ matrices.} Altogether, we may then conclude that $\hat{\T}$ converges point-wise to an ``infinite-size identity matrix'' which has the property that half of its eigenvalues equal zero. The convergence \eqref{eq:RntoR} therefore clearly does not hold here as the spectral norm (or the absolute largest eigenvalue) of $\hat{\T}-\I_N$ is greater or equal to $1$ for all $N$. 

\begin{figure}
  \centering
  \begin{tikzpicture}[font=\footnotesize]
    \renewcommand{\axisdefaulttryminticks}{4} 
    %\pgfplotsset{every major grid/.append style={densely dashed}}       
    \tikzstyle{every axis y label}+=[yshift=-10pt] 
    \tikzstyle{every axis x label}+=[yshift=5pt]
    \pgfplotsset{every axis legend/.append style={cells={anchor=west},fill=white, at={(0.98,0.98)}, anchor=north east, font=\scriptsize }}
    \begin{axis}[
      %ybar,
      xmin=0,
      ymin=0,
      xmax=3,
      ymax=0.9,
      ytick={0,0.2,0.4,0.6,0.8},
      yticklabels = {$0$,$0.2$,$0.4$,$0.6$,$0.8$},
      bar width=5pt,
      grid=major,
      ymajorgrids=false,
      scaled ticks=true,
      %scale ticks above={4},
      xlabel={Eigenvalues of $\hat{\T}$},
      ylabel={Density}
      ]
      \addplot+[ybar,mark=none,color=black,fill=blue!40!white] coordinates{
      (0.050000,0.000000)(0.150000,0.000000)(0.250000,0.220000)(0.350000,0.800000)(0.450000,0.820000)(0.550000,0.840000)(0.650000,0.760000)(0.750000,0.780000)(0.850000,0.680000)(0.950000,0.620000)(1.050000,0.620000)(1.150000,0.560000)(1.250000,0.500000)(1.350000,0.480000)(1.450000,0.440000)(1.550000,0.380000)(1.650000,0.340000)(1.750000,0.320000)(1.850000,0.280000)(1.950000,0.220000)(2.050000,0.200000)(2.150000,0.120000)(2.250000,0.020000)(2.350000,0.000000)(2.450000,0.000000)(2.550000,0.000000)(2.650000,0.000000)(2.750000,0.000000)(2.850000,0.000000)(2.950000,0.000000)(3.050000,0.000000)(3.150000,0.000000)(3.250000,0.000000)(3.350000,0.000000)(3.450000,0.000000)(3.550000,0.000000)(3.650000,0.000000)(3.750000,0.000000)(3.850000,0.000000)(3.950000,0.000000)(4.050000,0.000000)(4.150000,0.000000)(4.250000,0.000000)(4.350000,0.000000)(4.450000,0.000000)(4.550000,0.000000)(4.650000,0.000000)(4.750000,0.000000)(4.850000,0.000000)(4.950000,0.000000)(5.050000,0.000000)
      };
      \addplot[smooth,red,line width=0.5pt] plot coordinates{
      (0.100000,0.000000)(0.200000,0.000000)(0.300000,0.662615)(0.400000,0.838401)(0.500000,0.842169)(0.600000,0.806315)(0.700000,0.759546)(0.800000,0.710650)(0.900000,0.662615)(1.000000,0.616404)(1.100000,0.572197)(1.200000,0.529853)(1.300000,0.489095)(1.400000,0.449584)(1.500000,0.410936)(1.600000,0.372721)(1.700000,0.334423)(1.800000,0.295379)(1.900000,0.254626)(2.000000,0.210542)(2.100000,0.159695)(2.200000,0.090357)(2.300000,0.000000)(2.400000,0.000000)(2.500000,0.000000)(2.600000,0.000000)(2.700000,0.000000)(2.800000,0.000000)(2.900000,0.000000)(3.000000,0.000000)
      };
      \legend{ {Empirical eigenvalues},{Mar\u{c}enko-Pastur density} }
    \end{axis}
  \end{tikzpicture}
  \caption{Histogram of the eigenvalues of a single realization of $\hat{\T}=\frac1n\sum_{k=1}^n\y_k\y_k^\herm$, $\y_k$ has proper complex Gaussian entries, for $n=2000$, $N=500$.}
  \label{fig:empiricalMPlaw}
\end{figure}
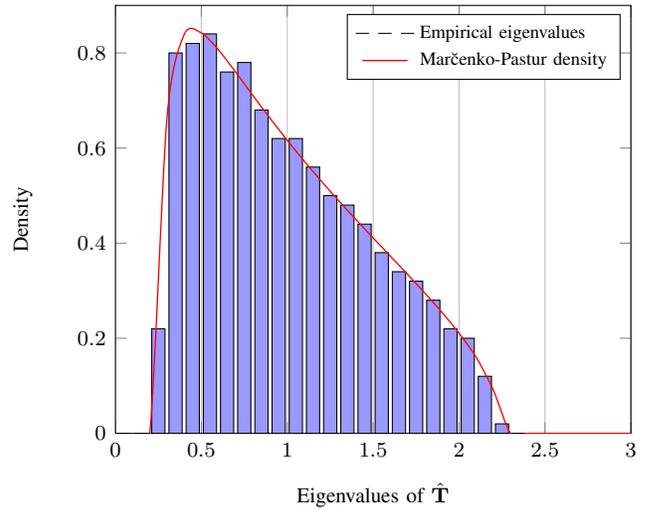

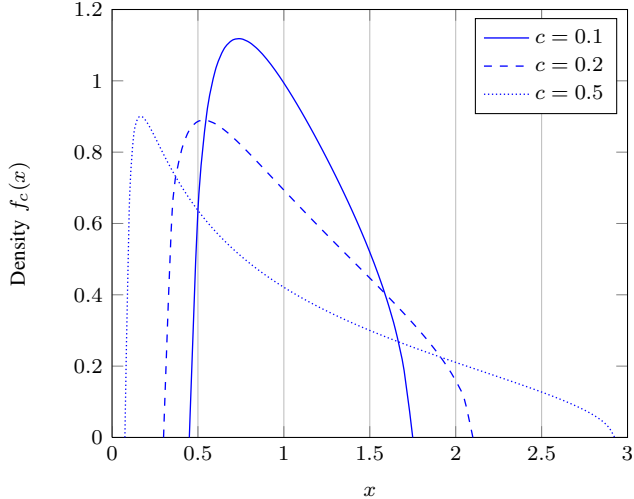
\begin{figure} 
  \centering
  \begin{tikzpicture}[font=\footnotesize]
    \renewcommand{\axisdefaulttryminticks}{8} 
    %\pgfplotsset{every major grid/.append style={densely dashed}}       
    \tikzstyle{every axis y label}+=[yshift=-10pt] 
    \tikzstyle{every axis x label}+=[yshift=5pt]

    \begin{axis}[
      grid=major,
      ymajorgrids=false,
      xlabel={$x$},
      ylabel={Density $f_c(x)$},
      xmin=0,
      xmax=3, 
      ymin=0, 
      ymax=1.2
      ]
      \addplot[smooth,blue,line width=0.5pt] plot coordinates{
(0.450000,0.000000) (0.500000,0.636620) (0.550000,0.903566) (0.600000,1.027341) (0.650000,1.088154) (0.700000,1.113853) (0.750000,1.117868) (0.800000,1.107672) (0.850000,1.087773) (0.900000,1.061033) (0.950000,1.029331) (1.000000,0.993922) (1.050000,0.955651) (1.100000,0.915077) (1.150000,0.872551) (1.200000,0.828269) (1.250000,0.782291) (1.300000,0.734561) (1.350000,0.684894) (1.400000,0.632955) (1.450000,0.578207) (1.500000,0.519798) (1.550000,0.456323) (1.600000,0.385253) (1.650000,0.301189) (1.700000,0.187241) (1.750000,0.000000)
      };
      \addplot[smooth,blue,dashed,line width=0.5pt] plot coordinates{
(0.300000,0.000000) (0.350000,0.632955) (0.400000,0.795775) (0.450000,0.861806) (0.500000,0.886137) (0.550000,0.888968) (0.600000,0.879762) (0.650000,0.863522) (0.700000,0.843089) (0.750000,0.820159) (0.800000,0.795775) (0.850000,0.770594) (0.900000,0.745035) (0.950000,0.719362) (1.000000,0.693740) (1.050000,0.668269) (1.100000,0.643000) (1.150000,0.617956) (1.200000,0.593135) (1.250000,0.568520) (1.300000,0.544077) (1.350000,0.519764) (1.400000,0.495529) (1.450000,0.471306) (1.500000,0.447021) (1.550000,0.422584) (1.600000,0.397887) (1.650000,0.372799) (1.700000,0.347154) (1.750000,0.320737) (1.800000,0.293254) (1.850000,0.264288) (1.900000,0.233194) (1.950000,0.198878) (2.000000,0.159155) (2.050000,0.108066) (2.100000,0.000000)
      };
      \addplot[smooth,blue,densely dotted,line width=0.5pt] plot coordinates{
(0.075000,0.000000) (0.100000,0.636620) (0.125000,0.842169) (0.150000,0.894042) (0.175000,0.899167) (0.200000,0.886137) (0.225000,0.865608) (0.250000,0.842169) (0.275000,0.817958) (0.300000,0.794004) (0.325000,0.770803) (0.350000,0.748577) (0.375000,0.727408) (0.400000,0.707300) (0.425000,0.688223) (0.450000,0.670123) (0.475000,0.652943) (0.500000,0.636620) (0.525000,0.621092) (0.550000,0.606303) (0.575000,0.592197) (0.600000,0.578725) (0.625000,0.565840) (0.650000,0.553499) (0.675000,0.541665) (0.700000,0.530300) (0.725000,0.519373) (0.750000,0.508854) (0.775000,0.498715) (0.800000,0.488932) (0.825000,0.479482) (0.850000,0.470344) (0.875000,0.461499) (0.900000,0.452928) (0.925000,0.444616) (0.950000,0.436547) (0.975000,0.428708) (1.000000,0.421084) (1.025000,0.413665) (1.050000,0.406439) (1.075000,0.399395) (1.100000,0.392524) (1.125000,0.385817) (1.150000,0.379265) (1.175000,0.372860) (1.200000,0.366594) (1.225000,0.360462) (1.250000,0.354455) (1.275000,0.348568) (1.300000,0.342795) (1.325000,0.337131) (1.350000,0.331570) (1.375000,0.326106) (1.400000,0.320737) (1.425000,0.315456) (1.450000,0.310260) (1.475000,0.305144) (1.500000,0.300105) (1.525000,0.295140) (1.550000,0.290243) (1.575000,0.285412) (1.600000,0.280645) (1.625000,0.275936) (1.650000,0.271284) (1.675000,0.266686) (1.700000,0.262138) (1.725000,0.257637) (1.750000,0.253182) (1.775000,0.248769) (1.800000,0.244396) (1.825000,0.240060) (1.850000,0.235759) (1.875000,0.231490) (1.900000,0.227251) (1.925000,0.223039) (1.950000,0.218852) (1.975000,0.214687) (2.000000,0.210542) (2.025000,0.206415) (2.050000,0.202302) (2.075000,0.198202) (2.100000,0.194112) (2.125000,0.190029) (2.150000,0.185950) (2.175000,0.181873) (2.200000,0.177794) (2.225000,0.173710) (2.250000,0.169618) (2.275000,0.165514) (2.300000,0.161396) (2.325000,0.157257) (2.350000,0.153096) (2.375000,0.148905) (2.400000,0.144681) (2.425000,0.140418) (2.450000,0.136109) (2.475000,0.131747) (2.500000,0.127324) (2.525000,0.122831) (2.550000,0.118257) (2.575000,0.113590) (2.600000,0.108815) (2.625000,0.103915) (2.650000,0.098869) (2.675000,0.093649) (2.700000,0.088223) (2.725000,0.082546) (2.750000,0.076561) (2.775000,0.070184) (2.800000,0.063296) (2.825000,0.055701) (2.850000,0.047055) (2.875000,0.036616) (2.900000,0.021952) (2.925000,0.000000) 
      };
      \legend{ {$c=0.1$},{$c=0.2$},{$c=0.5$} }
    \end{axis}
  \end{tikzpicture}
  \caption{Mar\u{c}enko-Pastur density $f_c$ for different limit ratios $c=\lim N/n$.}
  \label{fig:MPlaw}
\end{figure}

This convergence paradox (convergence of $\hat{\T}$ to an identity matrix with zero eigenvalues) arises obviously from an incorrect reasoning when taking the limits $N,n\to\infty$. Indeed, the convergence of matrices with increasing sizes only makes sense if a proper measure, here the spectral norm, in the space of such matrices is considered. The outcome of our previous observation is then that the {\it random} (or empirical) eigenvalue distribution $F^{\hat{\T}}$ of $\hat{\T}$, defined as
\begin{equation*}
	F^{\hat{\T}}(x) = \frac1N\sum_{i=1}^N 1_{x\leq \lambda_i}
\end{equation*}
with $\lambda_1,\ldots,\lambda_N$ the eigenvalues of $\hat{\T}$, does not converge (weakly) to a Dirac mass in $1$, and this is all what can be said at this point. This remark has fundamental consequences for basic signal processing detection and estimation procedures.  
In fact, although the eigenvalue distribution of $\hat{\T}$ does not converge to a Dirac in $1$, it turns out that in many situations of practical interest, it does converge towards a limiting distribution, as $N,n\to\infty$, $N/n\to c\geq 0$. In the particular case where $\T=\I_N$, the limiting distribution is known as the Mar\u{c}enko-Pastur law, initially studied by Mar\u{c}enko and Pastur in \cite{MAR67}. This distribution is depicted in Figure~\ref{fig:empiricalMPlaw} and compared to the empirical eigenvalue distribution of a matrix $\hat{\T}$ of large dimensions. From this figure, we observe that the empirical distribution of the eigenvalues is a good match to the limiting distribution for $N,n$ large. It is also observed that no eigenvalue seems to lie outside the support of the limiting distribution even for finite $N,n$. This fundamental property can be proved to hold for any matrix $\Y$ with independent zero mean and unit variance entries, along with some mild assumptions on the higher order moments. The shape of the Mar\u{c}enko-Pastur law for different limiting ratios $c$ is depicted in Figure~\ref{fig:MPlaw}. This figure suggests that, as $c\to 0$, the support of the Mar\u{c}enko-Pastur law tends to concentrate into a single mass in $1$. This is in line with our expectations from the discussion above for finite $N$, while the support tends to spread and eventually reaches $0$ when $c$ is large, which is compliant with the existence of a mass of eigenvalues at $0$ when $N>n$, i.e. $c> 1$. This will be confirmed by the explicit expression of the limiting density given later in Equation~\ref{eq:MPdensity}.

The general tools employed to derive the Mar\u{c}enko-Pastur law and its generalizations to some more advanced random matrix models are briefly introduced in Section~\ref{sec:rmt}. % as an example of use of one of the most fundamental tools of random matrix theory: the {\it Stieltjes transform}. 

The main implication of the above observations on signal processing methods is that, in general, the tools developed in view of applications for small $N$ and large $n$ are no longer adequate if either $N$ is taken much larger, or if $n$ cannot be afforded to be large. Observe for instance in Figure~\ref{fig:MPlaw} that, even for $n$ being ten times larger than $N$, the spread of the Mar\u{c}enko-Pastur law around $1$ is significant enough for the classical large $n$ assumption to be quite disputable. In the remainder of this section, we introduce a practical inference problem which assumes the setup $n\gg N$ and which can be proved asymptotically biased if $N/n$ does not converge to zero.  

\subsection{An eigenvalue inference problem}
\label{sec:multisource}
The following example deals with the statistical inference of eigenvalues of a large population covariance matrix. This problem is very generic and finds applications in array processing, such as radar detection, where the objective is to enumerate the sources and to infer their distances to the radar, or in cognitive radios where the objective is to detect and estimate the power of concurrent signal transmissions. This example will be used throughout the article to demonstrate the transition from the classical signal processing considerations towards the advanced random matrix approaches.

We assume that one has access to successive independent realizations $\y_1,\ldots,\y_n$ of a stationary process $\y_t=\U \x_t\in\CC^N$, with $\U\in\CC^{N\times N}$ an unknown unitary matrix and $\x_t\in\CC^N$ a complex circularly symmetric Gaussian vector with zero mean and covariance matrix $\P$. We further assume that $\P$ is diagonal and composed of $N_i$ eigenvalues equal to $P_i$, for $1\leq i\leq K$, with $P_1<\ldots<P_K$, and where $N=\sum_{i=1}^K N_i$. The presence of the unknown matrix $\U$ translates the fact that, in many applications, the eigenvector structure of the population covariance matrix $\T=\U\P\U^\herm$ of $\y_t$ is not necessarily known. Hence, the diagonal entries of $\P$ are not directly accessible. Note that this is in sharp contrast to covariance matrix estimation of structured signals, as in the recent works \cite{BIC08,WU09} where population covariance matrices have a Toeplitz-like structure. Here, no such assumption on $\U$ is made, so that alternative techniques need to be used in order to estimate the entries of $\P$. 

Our objective is to infer the values of $P_1,\ldots,P_K$ in the setting where both $n$ and $N$ are of similar order of magnitude or, stated in signal processing terms, when the number of observations of the process $\y_t$ is {\it not} large compared to the system dimension $N$. As before, for readability, we denote $\Y=[\y_1,\ldots,\y_n]$ and recall that $\frac1n\Y\Y^\herm$ is the sample covariance matrix of these data. 

Let us first consider the classical approach which assumes that $n\gg N$. We then have from a direct application of the law of large numbers that
\begin{equation}
	\label{eq:YYP}
\left\Vert \frac1n\Y\Y^\herm - \U\P\U^\herm\right\Vert \asto 0
\end{equation}
as $n\to\infty$. Since the convergence is in spectral norm and since the eigenvalues of $\U\P\U^\herm$ are the same as those of $\P$, an (asymptotically) {\it $n$-consistent} estimate $\hat{P}^{\infty}_k$ for $P_k$ reads
\begin{equation}
	\label{eq:Pinfty}
	\hat{P}^{\infty}_k = \frac1{N_k}\sum_{i\in\mathcal N_k} \lambda_i 
\end{equation}
where $\lambda_1\leq\ldots\leq\lambda_N$ are the ordered eigenvalues of $\frac1n\Y\Y^\herm$ gathered in successive clusters $\mathcal N_k=\{N-\sum_{i=k}^{K}N_i+1,\ldots,N-\sum_{i=k+1}^K N_i\}$. Indeed, the eigenvalues of $\frac1n\Y\Y^\herm$ indexed by $\mathcal N_k$ can be directly mapped to $P_k$ (their limiting value in the large $n$ regime). %, as can already be observed in Figure~\ref{fig:seperation}.

\begin{figure}
  \centering
  \begin{tikzpicture}[font=\footnotesize,scale=1]
    \renewcommand{\axisdefaulttryminticks}{4} 
    %\pgfplotsset{every major grid/.append style={densely dashed}}       
    \tikzstyle{every axis y label}+=[yshift=-10pt] 
    %\tikzstyle{every axis x label}+=[yshift=0pt]
    \pgfplotsset{every axis x label/.append style={below, yshift=5pt}}
    \pgfplotsset{every axis legend/.append style={cells={anchor=west},fill=white, at={(0.98,0.98)}, anchor=north east, font=\scriptsize }}
    \begin{axis}[
      %ybar,
      xmin=0,
      ymin=0,
      xmax=11,
      ymax=0.7,
      xtick={1,3,7},
      bar width=3pt,
      grid=major,
      ymajorgrids=false,
      scaled ticks=true,
      %scale ticks above={4},
      xlabel={Eigenvalues},
      ylabel={Density}
      ]
      \addplot+[ybar,mark=none,color=black,fill=blue!40!white] coordinates{
      (0.000000,0.000000)(0.200000,0.000000)(0.400000,0.000000)(0.600000,0.516667)(0.800000,0.616667)(1.000000,0.433333)(1.200000,0.100000)(1.400000,0.000000)(1.600000,0.000000)(1.800000,0.033333)(2.000000,0.133333)(2.200000,0.216667)(2.400000,0.183333)(2.600000,0.216667)(2.800000,0.200000)(3.000000,0.183333)(3.200000,0.183333)(3.400000,0.133333)(3.600000,0.116667)(3.800000,0.066667)(4.000000,0.000000)(4.200000,0.000000)(4.400000,0.000000)(4.600000,0.000000)(4.800000,0.000000)(5.000000,0.033333)(5.200000,0.083333)(5.400000,0.066667)(5.600000,0.066667)(5.800000,0.083333)(6.000000,0.083333)(6.200000,0.100000)(6.400000,0.083333)(6.600000,0.083333)(6.800000,0.100000)(7.000000,0.083333)(7.200000,0.083333)(7.400000,0.083333)(7.600000,0.066667)(7.800000,0.066667)(8.000000,0.083333)(8.200000,0.066667)(8.400000,0.083333)(8.600000,0.050000)(8.800000,0.050000)(9.000000,0.050000)(9.200000,0.050000)(9.400000,0.016667)(9.600000,0.033333)(9.800000,0.016667)(10.000000,0.000000)(10.200000,0.000000)(10.400000,0.000000)(10.600000,0.000000)(10.800000,0.000000)(11.000000,0.000000)
      };
      \addplot[smooth,red,line width=0.5pt] plot coordinates{
      (0.100000,0.000000)(0.200000,0.000000)(0.300000,0.000000)(0.400000,0.000000)(0.500000,0.000000)(0.600000,0.000000)(0.700000,0.566479)(0.800000,0.639909)(0.900000,0.621965)(1.000000,0.557297)(1.100000,0.457428)(1.200000,0.312185)(1.300000,0.000000)(1.400000,0.000000)(1.500000,0.000000)(1.600000,0.000000)(1.700000,0.000000)(1.800000,0.000000)(1.900000,0.000000)(2.000000,0.089903)(2.100000,0.153269)(2.200000,0.180234)(2.300000,0.196326)(2.400000,0.205748)(2.500000,0.210590)(2.600000,0.212068)(2.700000,0.210954)(2.800000,0.207764)(2.900000,0.202847)(3.000000,0.196439)(3.100000,0.188690)(3.200000,0.179682)(3.300000,0.169428)(3.400000,0.157870)(3.500000,0.144855)(3.600000,0.130088)(3.700000,0.113010)(3.800000,0.092467)(3.900000,0.065387)(4.000000,0.000000)(4.100000,0.000000)(4.200000,0.000000)(4.300000,0.000000)(4.400000,0.000000)(4.500000,0.000000)(4.600000,0.000000)(4.700000,0.000000)(4.800000,0.000000)(4.900000,0.000000)(5.000000,0.035640)(5.100000,0.044591)(5.200000,0.054893)(5.300000,0.062052)(5.400000,0.067616)(5.500000,0.072020)(5.600000,0.075561)(5.700000,0.078418)(5.800000,0.080717)(5.900000,0.082554)(6.000000,0.083998)(6.100000,0.085106)(6.200000,0.085921)(6.300000,0.086479)(6.400000,0.086809)(6.500000,0.086936)(6.600000,0.086880)(6.700000,0.086660)(6.800000,0.086287)(6.900000,0.085777)(7.000000,0.085140)(7.100000,0.084382)(7.200000,0.083519)(7.300000,0.082543)(7.400000,0.081481)(7.500000,0.080309)(7.600000,0.079069)(7.700000,0.077714)(7.800000,0.076310)(7.900000,0.074781)(8.000000,0.073215)(8.100000,0.071535)(8.200000,0.069761)(8.300000,0.067999)(8.400000,0.065979)(8.500000,0.063963)(8.600000,0.062001)(8.700000,0.059849)(8.800000,0.057469)(8.900000,0.055018)(9.000000,0.052616)(9.100000,0.050132)(9.200000,0.046261)(9.300000,0.044456)(9.400000,0.039696)(9.500000,0.036288)(9.600000,0.034067)(9.700000,0.030621)(9.800000,0.028210)(9.900000,0.018762)(10.000000,0.000000)(10.100000,0.000000)(10.200000,0.000000)(10.300000,0.000000)(10.400000,0.000000)(10.500000,0.000000)(10.600000,0.000000)(10.700000,0.000000)(10.800000,0.000000)(10.900000,0.000000)(11.000000,0.000000)
      };
      \legend{ {Empirical eigenvalue distribution},{Limit law (from Theorem~\ref{th:BaiSil95})} }
    \end{axis}
  \end{tikzpicture}
  \begin{tikzpicture}[font=\footnotesize,scale=1]
    \renewcommand{\axisdefaulttryminticks}{4} 
    %\pgfplotsset{every major grid/.append style={densely dashed}}       
    \pgfplotsset{every axis x label/.append style={below, yshift=5pt}}
    %\pgfplotsset{every axis y label/.append style={left, xshift=-5pt}}
    \pgfplotsset{every axis legend/.append style={cells={anchor=west},fill=white, at={(0.98,0.98)}, anchor=north east, font=\scriptsize }}
    \begin{axis}[
      %ybar,
      xmin=0,
      ymin=0,
      xmax=11,
      ymax=0.7,
      xtick={1,3,4},
      bar width=3pt,
      grid=major,
      ymajorgrids=false,
      scaled ticks=true,
      %scale ticks above={4},
      xlabel={Eigenvalues},
      ylabel={Density}
      ]
      \addplot+[ybar,mark=none,color=black,fill=blue!40!white] coordinates{
      (0.000000,0.000000)(0.200000,0.000000)(0.400000,0.000000)(0.600000,0.533333)(0.800000,0.616667)(1.000000,0.450000)(1.200000,0.066667)(1.400000,0.000000)(1.600000,0.000000)(1.800000,0.066667)(2.000000,0.200000)(2.200000,0.233333)(2.400000,0.233333)(2.600000,0.233333)(2.800000,0.250000)(3.000000,0.200000)(3.200000,0.216667)(3.400000,0.200000)(3.600000,0.200000)(3.800000,0.183333)(4.000000,0.150000)(4.200000,0.166667)(4.400000,0.166667)(4.600000,0.133333)(4.800000,0.133333)(5.000000,0.116667)(5.200000,0.083333)(5.400000,0.100000)(5.600000,0.050000)(5.800000,0.016667)(6.000000,0.000000)(6.200000,0.000000)(6.400000,0.000000)(6.600000,0.000000)(6.800000,0.000000)(7.000000,0.000000)(7.200000,0.000000)(7.400000,0.000000)(7.600000,0.000000)(7.800000,0.000000)(8.000000,0.000000)(8.200000,0.000000)(8.400000,0.000000)(8.600000,0.000000)(8.800000,0.000000)(9.000000,0.000000)(9.200000,0.000000)(9.400000,0.000000)(9.600000,0.000000)(9.800000,0.000000)(10.000000,0.000000)(10.200000,0.000000)(10.400000,0.000000)(10.600000,0.000000)(10.800000,0.000000)(11.000000,0.000000)
      };
      \addplot[smooth,red,line width=0.5pt] plot coordinates{
      (0.100000,0.000000)(0.200000,0.000000)(0.300000,0.000000)(0.400000,0.000000)(0.500000,0.000000)(0.600000,0.142153)(0.700000,0.576404)(0.800000,0.645208)(0.900000,0.624288)(1.000000,0.556701)(1.100000,0.452971)(1.200000,0.300019)(1.300000,0.000000)(1.400000,0.000000)(1.500000,0.000000)(1.600000,0.000000)(1.700000,0.000000)(1.800000,0.000000)(1.900000,0.076814)(2.000000,0.162113)(2.100000,0.196245)(2.200000,0.216823)(2.300000,0.229463)(2.400000,0.236829)(2.500000,0.240438)(2.600000,0.241251)(2.700000,0.239922)(2.800000,0.236923)(2.900000,0.232615)(3.000000,0.227305)(3.100000,0.221280)(3.200000,0.214850)(3.300000,0.208371)(3.400000,0.202215)(3.500000,0.196668)(3.600000,0.191803)(3.700000,0.187475)(3.800000,0.183447)(3.900000,0.179507)(4.000000,0.175505)(4.100000,0.171351)(4.200000,0.166990)(4.300000,0.162395)(4.400000,0.157546)(4.500000,0.152434)(4.600000,0.147046)(4.700000,0.141373)(4.800000,0.135396)(4.900000,0.129095)(5.000000,0.122437)(5.100000,0.115381)(5.200000,0.107864)(5.300000,0.099801)(5.400000,0.091060)(5.500000,0.081439)(5.600000,0.070550)(5.700000,0.057810)(5.800000,0.043351)(5.900000,0.014458)(6.000000,0.000000)(6.100000,0.000000)(6.200000,0.000000)(6.300000,0.000000)(6.400000,0.000000)(6.500000,0.000000)(6.600000,0.000000)(6.700000,0.000000)(6.800000,0.000000)(6.900000,0.000000)(7.000000,0.000000)(7.100000,0.000000)(7.200000,0.000000)(7.300000,0.000000)(7.400000,0.000000)(7.500000,0.000000)(7.600000,0.000000)(7.700000,0.000000)(7.800000,0.000000)(7.900000,0.000000)(8.000000,0.000000)(8.100000,0.000000)(8.200000,0.000000)(8.300000,0.000000)(8.400000,0.000000)(8.500000,0.000000)(8.600000,0.000000)(8.700000,0.000000)(8.800000,0.000000)(8.900000,0.000000)(9.000000,0.000000)(9.100000,0.000000)(9.200000,0.000000)(9.300000,0.000000)(9.400000,0.000000)(9.500000,0.000000)(9.600000,0.000000)(9.700000,0.000000)(9.800000,0.000000)(9.900000,0.000000)(10.000000,0.000000)(10.100000,0.000000)(10.200000,0.000000)(10.300000,0.000000)(10.400000,0.000000)(10.500000,0.000000)(10.600000,0.000000)(10.700000,0.000000)(10.800000,0.000000)(10.900000,0.000000)(11.000000,0.000000)
      };
      \legend{ {Empirical eigenvalue distribution},{Limit law (from Theorem~\ref{th:BaiSil95})} }
    \end{axis}
  \end{tikzpicture}
  \caption{Histogram of the eigenvalues of $\frac1n\Y\Y^\herm$ for $N=300$, $n=3000$, with $\P$ diagonal composed of three evenly weighted masses in (i) $1$, $3$ and $7$ at the top, (ii) $1$, $3$, and $4$ on the bottom.}
  \label{fig:seperation}
\end{figure}
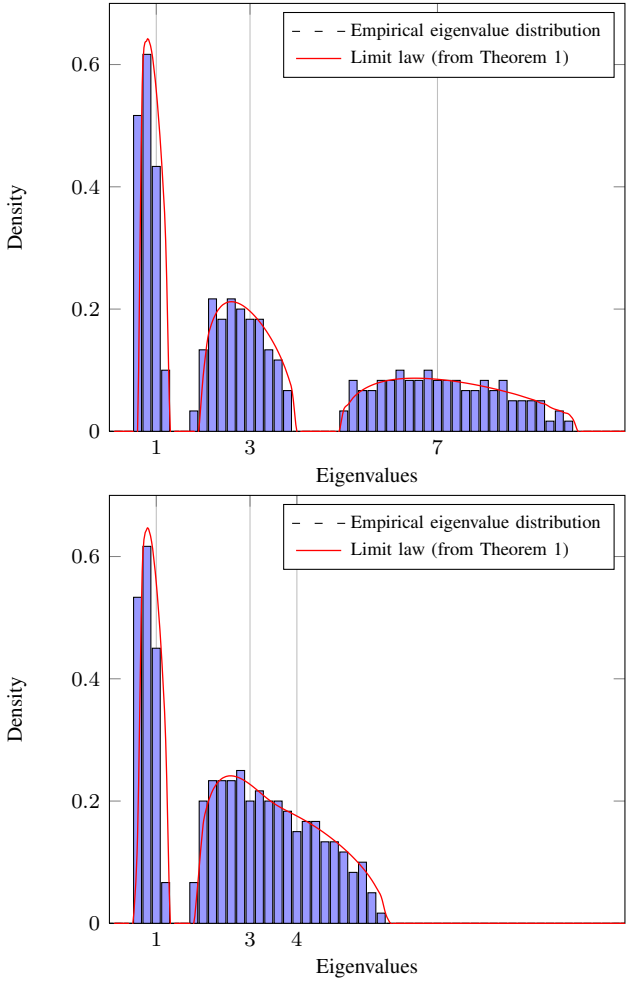

However, for $N$ and $n$ of comparable sizes, the convergence \eqref{eq:YYP} no longer holds. In this scenario, the spectrum of $\frac1n\Y\Y^\herm$ converges weakly to the union of a maximum of $K$ compact {\it clusters} of eigenvalues concentrated somewhat around $P_1,\ldots,P_K$, but whose centers of mass are {\it not} located at $P_1,\ldots,P_K$. This is depicted in Figure~\ref{fig:seperation}, where it is clearly seen that eigenvalues gather somewhat around the empirical values of $P_1,P_2,P_3$ in clusters. Therefore, as $N,n\to\infty$, the estimator $\hat{P}^\infty_k$ is no longer (asymptotically) consistent. Note also that, depending on the values of $P_1,\ldots,P_K$ (and in fact also on the ratio $N/n$), the number of clusters varies. From the second plot of Figure~\ref{fig:seperation}, we already anticipate the well-known problem of order selection, i.e. determining the number $K$ of distinct input sources from an observation $\Y$. For these reasons, more advanced analyses of the spectrum of $\frac1n\Y\Y^\herm$ must be carried out, from which {\it $(N,n)$-consistent} alternatives to \eqref{eq:Pinfty} can be derived, i.e. estimators which are consistent as both $N,n$ grow large. This introduction of a methodological approach to $(N,n)$-consistent estimators, called G-estimators, is one of the objectives of this article.

\section{Large dimensional random matrix theory}
\label{sec:rmt}

In this section, we introduce the basic notions of large dimensional random matrix theory and describe two elementary tools: the resolvent and the Stieltjes transform. These tools are necessary to understand the stark difference between the large random matrix techniques and the conventional statistical methods.

Random matrix theory studies the properties of matrices whose entries follow a joint probability distribution. The two main axes deal with the exact statistical characterization of small-size matrices and the asymptotic behaviour of large dimensional matrices. For the analysis of small-size matrices, one usually seeks closed form expressions of the exact statistics. However, these expressions become quickly intractable for practical analysis as one departs from white Gaussian matrices \cite{WIS28,JAM64,RAT05}. The mathematical methods used are usually very standard, mostly based on standard tools from probability theory. In the context of large dimensional random matrices, the tools are very different: asymptotic statistics and probability are obviously fundamental, but also linear algebra, as well as real and complex analysis, and combinatorics. The focus of these techniques are either based on invariance properties of the underlying matrices (free probability theory \cite{PET06,BIA03}, combinatorics \cite{RYA09b,RAO07}, Gaussian integration by part methods \cite{PAS00b,HAC06}) or on independence properties of their entries \cite{MAR67,SIL95,MES08b}. In the following, we will mostly concentrate on tools for large dimensional random matrices with independent entries, although some results we present originate from other techniques. The main tool of interest in the spectral study of these matrices is the {\it Stieltjes transform}, which we introduce hereafter.

\subsection{The Stieltjes transform}
Similar to the celebrated Fourier transform in classical probability theory and signal processing which allows one to perform simpler analysis in the Fourier (or frequency) domain than in the initial (or time) domain, the {\it spectral analysis} of large dimensional random matrices\footnote{What one refers here to as spectrum is the eigenvalue distribution of the underlying matrix.} is often carried out with the help of the Stieltjes transform.
\vspace{0.2cm}\begin{definition}[Stieltjes transform]
	\label{def:stieltjes}
Let $F$ be a real probability distribution function and $z\in\CC$ taken outside the support $\mathcal S$ of $F$. Then the (Cauchy-)Stieltjes transform $m_F(z)$ of $F$ at point $z$ is defined as
\begin{equation*}
	m_F(z) \triangleq \int_{\mathcal S} \frac1{t-z}dF(t).
\end{equation*}
\end{definition}\vspace{0.2cm}
Note importantly that, for $\X\in\CC^{N\times N}$ Hermitian with eigenvalues $\lambda_1,\ldots,\lambda_N$, and eigenvalue distribution $F^\X$, the Stieltjes transform $m_\X$ of $\X$ is, for $z\in \CC\setminus\{\lambda_1,\ldots,\lambda_N\}$,
\begin{align*}
	m_\X(z) = \int \frac{dF^\X(t)}{t-z} = \frac1N\sum_{k=1}^N \frac1{\lambda_k-z} = \frac1N\tr(\X-z\I_N)^{-1}.
\end{align*}
%This remark is essential when it comes to studying the properties of the eigenvalues of $\X$, as it is relatively easy to manipulate matrices of the type $(\X-z\I_N)^{-1}$ thanks to classical matrix inversion lemmas.
The Stieltjes transform, in the same way as the Fourier transform, has an inverse formula, given by
\begin{equation}
	\label{eq:inverseStieltjes}
	F(b)-F(a) = \lim_{y\downarrow 0} \int_a^b \Im\left[ m_F(x+\imath y)\right] dx.
\end{equation}
This formula is however rarely used in practice. What this says for random matrices is that, if the Stieltjes transform of a matrix $\X$ is known, then one can retrieve its eigenvalue distribution, which is often more difficult to obtain in the spectral domain than in the Stieltjes transform domain.

Since we will constantly deal with matrices of the form $(\X-z\I_N)^{-1}$, called the {\it resolvent} matrix of $\X$, the analysis of random matrices is fundamentally based on the exploitation of classical matrix inversion formulas. In order to better capture the essence of the Stieltjes transform approach, we hereafter outline the core arguments of the proof of the Mar\u{c}enko-Pastur law, as performed in the original article \cite{MAR67} back in 1967.

\subsection{The Mar\u{c}enko-Pastur law}
We recall that we wish to determine the expression of the limit of the eigenvalue distribution of $\Y\Y^\herm$, given in Figure~\ref{fig:MPlaw}, as $N,n\to\infty$, $N/n\to c$, where $\Y=[\y_1,\ldots,\y_n]\in\CC^{N\times n}$ has i.i.d. entries of zero mean and variance $1/n$. In the Stieltjes transform domain, we therefore need to evaluate $m_{\Y\Y^\herm}(z)=\frac1N\tr(\Y\Y^\herm-z\I_N)^{-1}$. 

The idea is to concentrate on the value of each individual diagonal entry of the resolvent $(\Y\Y^\herm-z\I_N)^{-1}$. Because of its obvious symmetrical structure, it suffices to study the asymptotic behaviour of the first diagonal entry. Writing $(\Y\Y^\herm-z\I_N)^{-1}$ as a block matrix with upper-left corner composed of the unique $(1,1)$ entry and applying the classical Schur complement \cite{HOR85}, it is easy to see that 
\begin{align}
	\label{eq:schur}
	\left[(\Y\Y^\herm-z\I_N)^{-1}\right]_{11} %&= \left[\begin{pmatrix} \tilde{\y}_1^\herm \tilde{\y}_1 -z & \tilde{\y}_1^\herm\tilde{\Y}^\herm \\ \tilde{\Y}\tilde{\y}_1 & \tilde{\Y}\tilde{\Y}^\herm - z\I_{N-1} \end{pmatrix}^{-1}\right]_{11} \nonumber\\
		&= \frac1{-z -z\tilde{\y}_1^\herm(\tilde{\Y}^\herm\tilde{\Y}-z\I_n)^{-1}\tilde{\y}_1}
\end{align}
where we have defined $\tilde{\Y}$ and $\tilde{\y}_1$ such that $\Y^\herm=[\tilde{\y}_1~\tilde{\Y}^\herm]$. Due to the independence between $\tilde{\y}_1$ and $\tilde{\Y}$ and the fact that $\tilde{\y}_1$ has i.i.d. entries, the quantity $\tilde{\y}_1^\herm(\tilde{\Y}^\herm\tilde{\Y}-z\I_n)^{-1}\tilde{\y}_1$ approaches $\frac1n\tr (\tilde{\Y}^\herm\tilde{\Y}-z\I_n)^{-1}$ as the system sizes $N,n$ grow large. Indeed, this is clearly true in the expectation over $\tilde{\y}_1$ (with equality) and, conditioned on $\tilde{\Y}$, the convergence somewhat recalls a law of large numbers. The interesting part is to notice that the latter trace expression is tightly connected to $\frac1N\tr (\Y\Y^\herm-z\I_N)^{-1}$, as the only major difference between the two lies in a single column change (a rank-one perturbation) which asymptotically will not alter the limiting normalized trace. A complete exposition of this derivation can be found in many random matrix books, e.g. \cite{SIL06,AND10,COUbook}.

Since this remark is valid for all diagonal entries of the resolvent $(\Y\Y^\herm-z\I_N)^{-1}$, it holds also true for their average $\frac1N\tr (\Y\Y^\herm-z\I_N)^{-1}$, i.e. for the Stieltjes transform $m_{\Y\Y^\herm}(z)$ of $\Y\Y^\herm$. Recalling \eqref{eq:schur}, the Stieltjes transform can therefore be asymptotically well approximated as a function of itself. This naturally leads $m_{\Y\Y^\herm}(z)$ to be approximately the solution of a {\it fixed-point equation} (or implicit equation). In many random matrix models, this is as far as the Stieltjes transform would go, and we would conclude that the limiting spectrum of the matrix under study is defined through its Stieltjes transform, whose expression is only known through a fixed-point equation (with usually a unique solution\footnote{For $z\in\RR^-$, the fixed-point map is usually a {\it standard interference function}, as described in \cite{YAT95}, allowing for simple algorithmic methods to solve the fixed-point in practice.}).

In the case of the Mar\u{c}enko-Pastur law, this fixed-point equation reads precisely
\begin{align} 
	\label{eq:fpm}
	m_{\Y\Y^\herm}(z) \simeq \frac1{1-c-z-zcm_{\Y\Y^\herm}(z)}
\end{align}
which turns out to be equivalent to a second order polynomial in $m_{\Y\Y^\herm}(z)$ which can be solved explicitly. Using the inverse Stieltjes transform formula \eqref{eq:inverseStieltjes}, we then obtain the limiting spectrum of $\Y\Y^\herm$ with density $f(x)$ given by
\begin{equation}
	\label{eq:MPdensity}
	f(x) = (1-c^{-1})^+ \delta(x) + \frac1{2\pi c x}\sqrt{(x-a)(b-x)}
\end{equation}
with $a=(1-\sqrt{c})^2$, $b=(1+\sqrt{c})^2$, and we recall that $c$ is the limiting ratio $\lim N/n$, where the square-root appearing in the expression originates from the roots of the second order polynomial defining $m_{\Y\Y^\herm}(z)$.

It is interesting to see that this expression is only defined for $a\leq x\leq b$ and possibly $x=0$, which defines the support of the distribution. In particular, we confirm, as already discussed that, as $c\to 0$, $[a,b]$ tends to a singleton in $1$. The convergence to $\{1\}$ is however quite slow as it is of order $\sqrt{c}$ for $c$ small (since $(1+\sqrt{c})^2\simeq 1+2\sqrt{c}$ in this regime). This suggests that $n$ must be much larger than $N$ for the classical large-$n$ approximation to be acceptable. Typically, in array processing with $100$ antennas, no less than $10\,000$ signal observations must be available for subspace methods relying on the large-$n$ approximation to be valid with an error of order $1/10$ on the assumed eigenvalues of $\Y\Y^\herm$. The resulting error may already lead to quite large fluctuations of the classical estimators. 

This observation strongly suggests that the classical techniques which assume $n\gg N$ must be revisited, as will be detailed in Section \ref{sec:eigeninference}. In the next section, we go a step further and introduce a generalization of the Mar\u{c}enko-Pastur law for more structured random matrices.

\subsection{Spectrum of large matrices}

As should have become clear from the discussion above, a central quantity of interest in many signal processing applications is the {\it sample covariance matrix} of $n$ realizations $\Y=[\y_1,\ldots,\y_n]$ of a process $\y_t = \T\x_t$ with $\x_t$ having i.i.d. zero mean and unit variance entries. The realizations may be independent, in which case $\Y=\T\X$, $\X=[\x_1,\ldots,\x_n]$ with i.i.d. entries, or have a (linear) time dependence structure, in which case $\y_t$ is often modelled as an autoregressive moving-average (ARMA) process; if so, we can write $\Y=\T\Z\R$ with $\Z$ a random matrix with i.i.d. zero mean entries and $\R$ a deterministic time correlation matrix (generally Toeplitz). The limiting eigenvalue distribution of many of such random matrix structures have been studied in the past ten years and have led to many fundamental results which only recently have found their way to signal processing applications. 

In the following, we treat the simplest case of a sample covariance matrix based on $n$ independent observations of the process $\y_t$ described above, when $\x_t$ has i.i.d. entries with zero mean. The outline of the techniques below is in general the same when it comes to more complicated models. 

We start with the important generalization of the Mar\u{c}enko-Pastur law to matrices with left-sided correlation, whose general proof is due to Bai and Silverstein \cite{SIL95}.
\vspace{0.2cm}\begin{theorem}
	\label{th:BaiSil95}
	Consider the matrix 
	\begin{equation*}
	\hat{\T} = \T^\oh\X\X^\herm\T^\oh=\sum_{i=1}^n \T^\oh\x_i\x_i^\herm\T^\oh
\end{equation*}
where $\X=[\x_1,\ldots,\x_n]\in \CC^{N\times n}$ has i.i.d. entries with zero mean and variance $1/n$, and $\T\in\CC^{N\times N}$ is nonnegative Hermitian whose eigenvalue distribution $F^{\T}$ converges weakly to $F^T$ as $N\to\infty$. Then, the eigenvalue distribution $F^{\hat{\T}}$ of $\hat{\T}$ converges weakly and almost surely to the distribution $F$ with Stieltjes transform $m_{F}(z)=cm_{\underline F}(z)+(c-1)\frac1z$, where $m_{\underline F}(z)$ is a solution to
\begin{align}
	\label{eq:Stiel_XTX_original}
	m_{\underline F}(z) &=-\left(z-c\int\frac{t}{1+tm_{\underline F}(z)}dF^T(t)\right)^{-1}
  \end{align}
  for all $z\in\CC^+\triangleq \{z\in\CC,~\Im[z]>0\}$.
\end{theorem}\vspace{0.2cm}

Note here that, contrary to the case treated previously where $\T=\I_N$, the limiting eigenvalue distribution of $\hat{\T}$ does not take an explicit form, but is only defined {\it implicitly} through its Stieltjes transform which satisfies a fixed-point equation. This result is nonetheless usually sufficient to derive {\it explicit} detection tests and estimators for our application needs. Although not mentioned for readability in the statement of the theorem, it is usually true that the fixed-point equation has a unique solution on some restricted set, and that classical fixed-point algorithms do converge surely to the proper solution; this is particularly convenient for numerical evaluations.

An important remark for the subsequent discussion is that the integral in \eqref{eq:Stiel_XTX_original} can be easily related to the Stieltjes transform of $m_T$ of $F^T$ to give the equivalent representation
\begin{align}
  \label{eq:Stiel_XTX}
  m_{\underline F}(z) &=-\left(z - \frac{c}{m_{\underline F}(z)} \left[1-\frac1{m_{\underline F}(z)} m_T\left(-\frac1{m_{\underline F}(z)}\right) \right] \right)^{-1}.
\end{align}
Although the original formula of \cite{SIL95} is given by \eqref{eq:Stiel_XTX_original}, Equation \eqref{eq:Stiel_XTX} is more interesting in several aspects. The important observation in \eqref{eq:Stiel_XTX} is that the Stieltjes transform $m_T$ can be related to the limiting Stieltjes transform $m_{\underline F}$ of $\hat{\T}$ through this simple equation. This is the fundamental anchor for {\it inference techniques}, where the {\it observable data} ($\hat{\T}$) get connected to the {\it hidden parameters} (information on $\T$) to be estimated. 

Indeed, in the sample covariance matrix model, eigen-inference methods used in subspace estimators consist in retrieving information about the population covariance matrix $\T$ from the sample covariance matrix $\hat{\T}$. Therefore, if one can relate the information to be estimated in $\T$ to $m_\T$ (which is asymptotically $m_T$), then \eqref{eq:Stiel_XTX} provides the fundamental link between $\T$ and $\hat{\T}$ through their respective limiting Stieltjes transform, from which estimators can be obtained. This procedure is detailed in the next section.

Note that Theorem~\ref{th:BaiSil95} provides the value of $m_F(z)$ for all $z\in\CC^+$. Therefore, if one desires to numerically evaluate the limiting eigenvalue distribution of $\hat{\T}$, it suffices to evaluate $m_{F}(z)$ for $z=x+\imath\varepsilon$, with $\varepsilon>0$ small and for all $x>0$ and then use the inverse Stieltjes transform formula \eqref{eq:inverseStieltjes} to describe the limiting spectrum density $f$ by
\begin{equation*}
	f(x) \simeq \frac1\pi \Im\left[m_F(x+\imath \varepsilon)\right].
\end{equation*}
We used this technique to produce Figure~\ref{fig:seperation} in which $\T=\U\P\U^\herm$ was assumed composed of three evenly weighted masses in $\{1,3,7\}$ (top) or $\{1,3,4\}$ (bottom).

As mentioned above, it is possible to derive the limiting eigenvalue distribution of a wide range of random matrix models or, if a limit does not exist for the model (this may arise for instance if the eigenvalues of $\T$ are not assumed to converge in law), to derive {\it deterministic equivalents} for the large dimensional matrices. For the latter, the eigenvalue distribution $F^{\hat{\T}}$ of $\hat{\T}$ can be approximated for each $N$ by a {\it deterministic} distribution $F_N$ such that $F^{\hat{\T}}-F_N\Rightarrow 0$ almost surely. These distribution functions $F_N$ usually have a very practical form for analysis; see e.g. \cite{DUP09,COU09,HAC07} for examples of random matrix models that do not admit limits but only deterministic equivalents. 

The knowledge of the limiting eigenvalue distribution of such matrices is not of particular interest to signal processing, but it is a required prior step before performing statistical inference and detection methods. In wireless communication problems however, such as that of evaluating the capacity 
\begin{align*}
	C&=\frac1N\log\det\left(\I_N+\frac1{\sigma^2} \H\H^\herm\right) \\
	&=\int_{\sigma^2}^\infty \left( \frac1t - \frac1N\tr(\H\H^\herm + t \I_N)^{-1} \right)dt
\end{align*}
of a multi-antenna Gaussian channel $\H$ with noise variance $\sigma^2$, these tools alone are fundamental, see e.g. \cite{HAC07,COU09,COU11,COU11d,DUP10}. In particular, we recognize in the above formula that the Stieltjes transform $\frac1N\tr(\H\H^\herm - z \I_N)^{-1}$ of $\H\H^\herm$, and not its eigenvalue distribution, is the quantity of central interest. This is probably the main explanation why wireless communication applications of random matrix results have appeared as early as in 1999 \cite{TSE99,SHA99}, while applications for signal processing methods emerged only very recently.

In the following, we pursue our study of the spectral properties of large dimensional random matrices to applications in statistical inference and introduce the eigen-inference (or G-estimation) method.

\subsection{G-estimation}
\label{sec:eigeninference}
As introduced in the previous section, one of the major objectives of random matrix theory for signal processing is to improve the statistical tests and inference techniques derived under the assumption of infinitely many observations for scenarios where the system population size $N$ and the number of observations $n$ are of the same order of magnitude. For this, we will rely on the connections between the Stieltjes transforms of the population and sample covariance matrices. 

The G-estimation method, originally due to Girko \cite{GIR90}, intends to provide such $(N,n)$-consistent estimators. The general technique consists in a three-step approach along the following lines. First, the parameter to be estimated, call it $\theta$, is written under the form of a functional of the Stieltjes transform of the deterministic matrices of the model, say here $\theta=f(m_{\T})$ for a population matrix $\T$ with Stieltjes transform $m_\T$. Then, $m_\T$ is connected to the Stieltjes transform $m_{\hat{\T}}$ of an observed matrix $\hat{\T}$ through a formula similar to that described earlier in \eqref{eq:Stiel_XTX}, e.g. $m_T=g(m_{\uF})$ for a certain function $g$. Finally, connecting the pieces together, we have a link between $\theta$ and the observations $\hat{\T}$, this link being only exact asymptotically. For finite $N,n$ dimensions, this naturally produces an (asymptotically) $(N,n)$-consistent estimator $\hat{\theta}=g(m_{\hat{\T}})$. 

While the fundamental tool for connecting population and observation spaces is the Stieltjes transform, the second important tool, that will help connecting the parameter $\theta$ to the Stieltjes transform $m_\T$, is the Cauchy integral formula and complex analysis in general.

For better understanding, we hereafter elaborate on a concrete example to describe the eigen-inference framework within the context of the estimation of the source powers discussed in Section~\ref{sec:multisource}. 

We recall that one observes a matrix of the type $\Y=\U\P^\oh\X$, where $\U\in\CC^{N\times N}$ is unitary, $\X\in\CC^{N\times n}$ is filled with i.i.d. entries with zero mean and unit variance and $\P$ is diagonal with $N_1$ entries equal to $P_1$, \ldots, and $N_K$ entries equal to $P_K$, all values of $P_i$ being distinct. 

While the objective in the previous section was to characterize the asymptotic eigenvalue distribution of $\frac1n\Y\Y^\herm$ when $\P$ is known, the target of this section is instead to infer $P_1,\ldots,P_K$ from the observation $\Y$. For this, we make the following fundamental observation. From Cauchy's complex integration formula \cite{RUD86},
\begin{equation*}
	P_k = -\frac1{2\pi \imath}\oint_{\mathcal C_k} \frac{z}{P_k-z}dz
\end{equation*}
for a complex positively oriented contour $\mathcal C_k$ circling once around $P_k$. If $\mathcal C_k$ does not enclose any of the other $P_i$, $i\neq k$, then it is also true, again from Cauchy's integration formula, that
\begin{align}
	P_k &= - \frac1{2\pi \imath}\frac{N}{N_k}\oint_{\mathcal C_k} z \frac1N\sum_{i=0}^K N_i\frac1{P_i-z}dz \nonumber \\ 
	\label{eq:PkmP}
	&= -\frac1{2\pi \imath}\frac{N}{N_k} \oint_{\mathcal C_k} z m_{\P}(z)dz.
\end{align}
Hence, we are able to write $P_k$ as a function of the Stieltjes transform of the non-observable matrix $\P$. The next step is to link $m_{\P}$ to the Stieltjes transform of $\frac1n\Y\Y^\herm$. For this, we use \eqref{eq:Stiel_XTX} with $\T=\U\P\U^\herm$ (which has the same eigenvalue spectrum as $\P$) and $\uF$ the limiting spectrum of $\frac1N\Y^\herm\Y$, which after the variable change $z=m_{\uF}(u)$ leads to the elementary expression
\begin{equation*}
	P_k = \frac1{c}\frac1{2\pi \imath}\oint_{\mathcal C'_k} u\frac{m_{\uF}'(u)}{m_{\uF}(u)}du
\end{equation*} 
for some contour $\mathcal C_k'$ which happens to circle exactly around the cluster of eigenvalues of $\hat{\T}$ associated to $P_k$ only, and with $m_{\uF}'$ the complex derivative of $m_{\uF}(z)$ with respect to $z$. It is therefore fundamental that there exists a contour that circles around the cluster associated to $P_k$ only. It may happen, depending on the ratio $c$ and the values of the $P_k$, that the limiting spectrum of $\hat{\T}$ generates less than $K$ clusters, some clusters being associated with multiple power values, as depicted in the second graph of Figure~\ref{fig:seperation}. In this scenario, the above result {\it does not} hold and the technique collapses. In signal processing applications, this effect is linked to the problem of {\it source separation}. From now on, we therefore assume that the cluster of eigenvalues associated to $P_k$ is perfectly isolated. Necessary and sufficient conditions for this to hold are clearly established \cite{MES08}.

Under the cluster separability assumption, an estimator $\hat{P}_k$ for $P_k$ is now straightforward to obtain. Indeed, replacing $m_{\uF}$ by its empirical estimate $m_{\frac1N\Y^\herm\Y}$, $P_k$ is approximately equal to a complex integral whose integrand is constituted of rational functions. A final step of residue calculus \cite{RUD86} completes the calculus and we obtain explicitly the estimator
\begin{equation}
	\label{eq:hatPk}
	\hat{P}_k = \frac{n}{N_k}\sum_{m\in\mathcal N_k}(\lambda_m-\mu_m)
\end{equation}
where $\mathcal N_k$ is defined as in \eqref{eq:Pinfty}, $\lambda_1\leq\ldots\leq\lambda_N$ are the eigenvalues of $\frac1n\Y\Y^\herm$, and $\mu_1\leq\ldots\leq\mu_N$ are the ordered eigenvalues of $\diag({\bm\lambda})-\frac1n\sqrt{\bm\lambda}\sqrt{\bm\lambda}^\trans$, with ${\bm\lambda}=(\lambda_1,\ldots,\lambda_N)^\trans$. This estimator therefore improves the $n$-consistent estimator \eqref{eq:Pinfty} to an $(N,n)$-consistent estimator which surprisingly turns out to have much better performance statistics for all couples $(N,n)$. Performance figures will be provided in Section~\ref{sec:applications} for signal processing scenarios of more practical interest.

Multiple estimators can be derived similarly for different types of models and problems, a large number of which have been investigated by Girko and named the G-estimators. A list of more than fifty estimators for various applications can be found in \cite{GIR00}. In Section~\ref{sec:applications}, some examples in the context of path loss estimation in wireless communications and angle of arrival estimation for array processing will be discussed.

We now turn to a finer analysis of random matrices which is no longer concerned with the weak limit of eigenvalue distributions but rather with the limiting fluctuations of individual eigenvalues (in particular the extreme eigenvalues). This problem is at the core of some standard statistical tests for signal detection, such as the generalized likelihood ratio test (GLRT).

\subsection{Extreme eigenvalues}
\label{sec:extreme}
It is important to note at this point that Theorem~\ref{th:BaiSil95} and all theorems deriving from the Stieltjes transform method only provide the weak convergence of the empirical eigenvalue distribution to a limit $F$ but do not say anything about the behaviour of the individual eigenvalues. In particular, the fact that the support of $F$ is compact does not mean that the empirical eigenvalues will asymptotically {\it all} lie in this compact support. Indeed, consider for instance the distribution $F_N(x)=\frac{N-1}N 1_{x\leq 1} + \frac1N 1_{x\leq 2}$. Then the largest value taken by a random variable distributed as $F_N$ equals $2$ for all $N$ although it is clear that $F_N$ converges weakly to $F=1_{x\leq 1}$ (and then the support of $F$ is the singleton $\{1\}$). In the context of Theorem~\ref{th:BaiSil95}, this reasoning indicates that, although the weak limit of $F^{\hat{\T}}$ is compactly supported, some or even many eigenvalues of $F^{\hat{\T}}$ may still be found outside the support.

As a matter of fact, it is proved in \cite{YIN88b} that, if the entries of $\X$ in Theorem~\ref{th:BaiSil95} have infinite fourth order moment, then the largest eigenvalue of $\X\X^\herm$ tends to infinity. For non-degenerate scenarios though, i.e. when the fourth moment of the entries of $\X$ is finite, we have the following much expected result \cite{YIN88,SIL98}.

\vspace{0.2cm}\begin{theorem}[No eigenvalue outside the support]
	\label{th:noeig}
	Let $\hat{\T}$ be defined as in Theorem~\ref{th:BaiSil95} with $\X$ having entries with finite fourth order moment and with $F^\T\Rightarrow F^T$, such that $\T$ has no eigenvalue outside the support of $F^T$. We recall that $F^{\hat{\T}}\Rightarrow F$ for $F$ defined in Theorem~\ref{th:BaiSil95}. Take now $[a,b]\subset \RR\cup\{\pm \infty\}$ strictly away from the support of $F$. Then, almost surely, there is no eigenvalue of $\hat{\T}$ found in $[a,b]$ for all large $N$.
\end{theorem}\vspace{0.2cm}

This result says in particular that, for $\T=\I_N$, and for all large $N$, there is no eigenvalue of $\hat{\T}$ found away from the support of the Mar\u{c}enko-Pastur law. This further generalizes to the scenario where $\T$ has a few distinct eigenvalues, each with large multiplicities, in which case $F$ is formed of multiple compact clusters, as depicted in Figure~\ref{fig:seperation}. In this context, it is proved in \cite{BAI99} that the number of eigenvalues asymptotically found in each cluster matches exactly the multiplicity of the corresponding mass in $\T$ (assuming cluster separability). This is often referred to as the property of {\it exact spectrum separation}. Note that this point fully justifies the last step in the derivation of the estimator \eqref{eq:hatPk} where we somewhat hid the fact that the residue calculus naturally imposes to know where the eigenvalues of $\hat{\T}$ are precisely located. As we will see in Section~\ref{sec:applications}, these results on the asymptotic location of the eigenvalues allow us to derive new hypothesis tests for the detection of signals embedded in white noise. In particular, we will present tests on the received sample covariance matrix $\hat{\T}$ that confront a pure white noise against a signal-plus-noise hypotheses, i.e. $\T=\I_N$ against $\T\neq \I_N$.

Based on the earlier reasoning on the weak limit interpretations, we insist that, while Theorem~\ref{th:BaiSil95} does state that $F^{\T}\Rightarrow 1_{x\leq 1}$ implies that $F$ is the Mar\u{c}enko-Pastur law, it does {\it not} state that $F^{\T}\Rightarrow 1_{x\leq 1}$ implies that no eigenvalue is asymptotically found away from the support of the Mar\u{c}enko-Pastur law. Indeed, if $\T=\diag(1,\ldots,1,a)$, with $a\neq 1$, the theorem cannot be applied. Conversely, for $\T$ described above, it is not clear whether an eigenvalue of $\hat{\T}$ will be seen outside the limiting support. This interrogation is of fundamental importance for the performance evaluation of asymptotic detection tests and has triggered a recent interest for this particular model for $\T$, called the {\it spike model}, which will be discussed next. 

\subsection{Spike models}
\label{sec:spike}
It is of particular interest for signal processing applications to model low rank signal spaces embedded in white noise. This is the case in array processing where a single signal source with unique propagation path is received during several consecutive time instants, hence producing a rank-one signal matrix to be added to the ambient full rank white noise. The objective of this particular model is to derive simple detection and identification procedures that do not fall in the rather involved scheme of Section~\ref{sec:eigeninference}. 

Under the generic denomination of ``spike models'' we understand all random matrix models showing small rank perturbations of some classical random matrix models. For instance, for $\X\in\CC^{N\times n}$ with i.i.d. entries with zero mean and variance $1/n$, the matrices $\hat{\T}=(\X+\E)(\X+\E)^\herm$, $\E\in\CC^{N\times n}$ such that $\E\E^\herm$ is of small rank $K$, or $\hat{\T}=(\I_N+\E)^\oh\X\X^\herm(\I_N+\E)^\oh$, $\E\in\CC^{N\times N}$ of small rank $K$, fall into the generic scheme of spike models. For all of these models, Theorem~\ref{th:BaiSil95} and similar results, see e.g. \cite{DOZ07}, claim that $F^{\hat{\T}}$ converges weakly to the Mar\u{c}enko-Pastur law, which is the major motivation of these models as they are rather simple to analyze. The interest here though is to study the behaviour of the extreme eigenvalues of $\hat{\T}$.

%The first remark concerning these spike models is that, to each eigenvalue (or singular value) of $\E$ corresponds an eigenvalue of $\hat{\T}$ asymptotically found either inside or outside the support of the Mar\u{c}enko-Pastur law. The condition for this eigenvalue to be found inside or outside the support of the limiting law depends on the amplitude of the eigenvalue of $\E$ under consideration and on the limiting ratio $c$. There therefore exists an asymptotic {\it phase transition} effect which rules the fact that remote eigenvalues of $\hat{\T}$ are found or not away from the base support, for all large $N$. This has important consequences related to decidability in large dimensional hypothesis tests.

To carry on with the same models as before, we concentrate here on $\hat{\T}$ modeled as in Theorem~\ref{th:BaiSil95}, with $\T$ taken to be a small rank perturbation of the identity matrix. Similar properties can be found for other models e.g. in \cite{BEN09,BAI08b}. The first result deals with first order limits of eigenvalues and eigenvector projections of the spike model \cite{BAI06,BEN09,COU11c}.
\vspace{0.2cm}\begin{theorem}
	\label{th:spike1}
	Let $\hat{\T}$ be defined as in Theorem~\ref{th:BaiSil95}, $\X$ have i.i.d. entries with zero mean and variance $1/n$, and $\T=\I_N+\E$ with $\E=\sum_{i=1}^K\omega_i\uu_i\uu_i^\herm$ its spectral decomposition (i.e. $\uu_i^\herm\uu_j=\delta_i^j$), where $K$ is fixed and $\omega_1>\dots>\omega_K>1$. Denote $\lambda_1\geq\ldots\geq\lambda_N$ the ordered eigenvalues of $\hat{\T}$ and $\hat{\uu}_{i}\in\CC^N$ the eigenvector associated with the eigenvalue $\lambda_i$. Then, for $1\leq i\leq K$, as $N,n\to\infty$ with $N/n\to c$,
	\begin{align*}
\lambda_i \asto \left\{
\begin{array}{ll}
	(1+\sqrt{c})^2 &,\textmd{ if }\omega_i\leq\sqrt{c} \\
	\rho_i \triangleq 1+\omega_i + c\frac{1+\omega_i}{\omega_i} &,\textmd{ if }\omega_i>\sqrt{c}
\end{array}
\right.
	\end{align*}
	and
	\begin{align*}
		|\uu_i^\herm \hat{\uu}_{i}| \asto \left\{
\begin{array}{ll}
	0 &,\textmd{ if }\omega_i\leq\sqrt{c} \\
	\xi_i \triangleq \frac{1-c\omega_i^{-2}}{1+c\omega_i^{-1}} &,\textmd{ if }\omega_i>\sqrt{c}.
\end{array}
\right.
	\end{align*}
\end{theorem}\vspace{0.2cm}

The main observation of this result is that, if $\omega_1\leq \sqrt{c}$, then the largest eigenvalue of $\hat{\T}$ asymptotically converges to the right edge of the Mar\u{c}enko-Pastur law and is therefore {\it hidden} in the main bulk of the eigenvalues, while if $\omega_1>\sqrt{c}$, the largest eigenvalue of $\hat{\T}$ is found outside the main cluster. This has fundamental consequences. When $\hat{\T}$ is the sample covariance matrix of signal-plus-noise data with signal strength $\omega_1$, then, if the signal is strong enough or conversely if $N/n$ is small enough, it is possible to detect this signal based on the presence of an eigenvalue exterior to the main eigenvalue cluster of $\hat{\T}$. Otherwise, neither the eigenvalues nor the eigenvectors of $\hat{\T}$ provide any information on the presence of a signal, at least asymptotically. As a conclusion, in the latter scenario, there is no way to decide if a signal is indeed present or not. This provides {\it fundamental limits} of signal detection tests. This subject is further discussed in the application Section~\ref{sec:applications}.

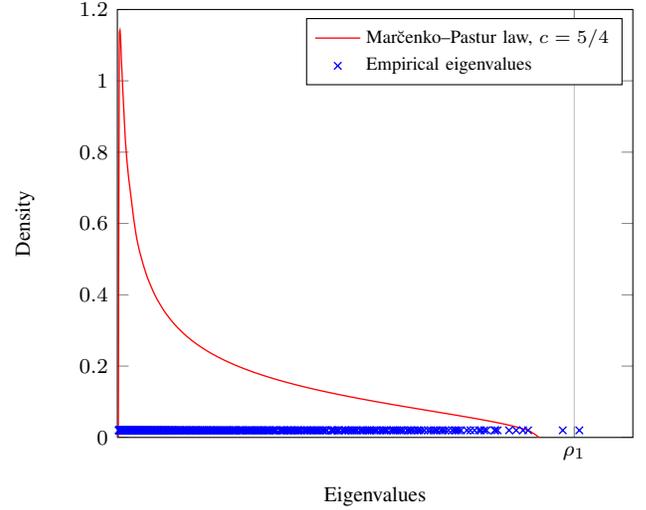
\begin{figure}
  \centering
  \begin{tikzpicture}[font=\footnotesize,scale=1]
    \renewcommand{\axisdefaulttryminticks}{4} 
    %\pgfplotsset{every major grid/.append style={densely dashed}}       
    \tikzstyle{every axis y label}+=[yshift=-10pt] 
    \tikzstyle{every axis x label}+=[yshift=5pt]
    \pgfplotsset{every axis legend/.append style={cells={anchor=west},fill=white, at={(0.98,0.98)}, anchor=north east, font=\scriptsize }}
    \begin{axis}[
      %ybar,
      xmin=0,
      ymin=0,
      xmax=5.5,
      ymax=1.2,
      xtick={4.875},
      xticklabels = {$\rho_1$},
      bar width=3pt,
      grid=major,
      ymajorgrids=false,
      scaled ticks=true,
      %scale ticks above={4},
      xlabel={Eigenvalues},
      ylabel={Density}
      ]
      \addplot[smooth,red,line width=0.5pt] plot coordinates{
      (0.010000,0.000000)(0.020000,1.048008)(0.030000,1.135652)(0.040000,1.083657)(0.050000,1.018592)(0.100000,0.782291)(0.200000,0.568520)(0.300000,0.464437)(0.400000,0.399793)(0.500000,0.354455)(0.600000,0.320249)(0.700000,0.293150)(0.800000,0.270914)(0.900000,0.252180)(1.000000,0.236065)(1.100000,0.221970)(1.200000,0.209470)(1.300000,0.198256)(1.400000,0.188095)(1.500000,0.178808)(1.600000,0.170257)(1.700000,0.162328)(1.800000,0.154933)(1.900000,0.147998)(2.000000,0.141460)(2.100000,0.135268)(2.200000,0.129379)(2.300000,0.123754)(2.400000,0.118360)(2.500000,0.113168)(2.600000,0.108152)(2.700000,0.103289)(2.800000,0.098557)(2.900000,0.093935)(3.000000,0.089404)(3.100000,0.084946)(3.200000,0.080541)(3.300000,0.076171)(3.400000,0.071814)(3.500000,0.067447)(3.600000,0.063045)(3.700000,0.058576)(3.800000,0.054001)(3.900000,0.049269)(4.000000,0.044307)(4.100000,0.039004)(4.200000,0.033174)(4.300000,0.026443)(4.400000,0.017779)(4.500000,0.000000)
      };
      \addplot[only marks,blue,mark=x,line width=0.5pt] plot coordinates{
      (4.929,0.02)(4.752,0.02)(4.382,0.02)(4.328,0.02)(4.255,0.02)(4.179,0.02)(4.057,0.02)(4.033,0.02)(4.024,0.02)(3.939,0.02)(3.914,0.02)(3.879,0.02)(3.858,0.02)(3.811,0.02)(3.766,0.02)(3.729,0.02)(3.666,0.02)(3.640,0.02)(3.617,0.02)(3.601,0.02)(3.574,0.02)(3.532,0.02)(3.498,0.02)(3.473,0.02)(3.449,0.02)(3.413,0.02)(3.396,0.02)(3.371,0.02)(3.343,0.02)(3.315,0.02)(3.274,0.02)(3.242,0.02)(3.218,0.02)(3.199,0.02)(3.168,0.02)(3.138,0.02)(3.131,0.02)(3.108,0.02)(3.093,0.02)(3.053,0.02)(3.029,0.02)(3.010,0.02)(2.988,0.02)(2.970,0.02)(2.950,0.02)(2.940,0.02)(2.917,0.02)(2.878,0.02)(2.863,0.02)(2.846,0.02)(2.826,0.02)(2.806,0.02)(2.771,0.02)(2.761,0.02)(2.739,0.02)(2.716,0.02)(2.697,0.02)(2.684,0.02)(2.667,0.02)(2.645,0.02)(2.464,0.02)(2.471,0.02)(2.623,0.02)(2.605,0.02)(2.590,0.02)(2.499,0.02)(2.507,0.02)(2.527,0.02)(2.550,0.02)(2.557,0.02)(2.112,0.02)(2.438,0.02)(2.433,0.02)(2.421,0.02)(2.398,0.02)(2.132,0.02)(2.148,0.02)(2.159,0.02)(2.185,0.02)(2.192,0.02)(2.203,0.02)(2.225,0.02)(2.236,0.02)(2.365,0.02)(2.359,0.02)(2.342,0.02)(2.317,0.02)(2.308,0.02)(2.280,0.02)(2.290,0.02)(2.287,0.02)(2.101,0.02)(1.887,0.02)(2.089,0.02)(2.073,0.02)(2.060,0.02)(2.050,0.02)(2.026,0.02)(2.021,0.02)(1.900,0.02)(2.001,0.02)(1.992,0.02)(1.916,0.02)(1.965,0.02)(1.954,0.02)(1.927,0.02)(1.936,0.02)(1.733,0.02)(1.744,0.02)(1.750,0.02)(1.765,0.02)(1.775,0.02)(1.785,0.02)(1.803,0.02)(1.874,0.02)(1.818,0.02)(1.826,0.02)(1.858,0.02)(1.843,0.02)(1.851,0.02)(1.567,0.02)(1.698,0.02)(1.694,0.02)(1.686,0.02)(1.582,0.02)(1.586,0.02)(1.669,0.02)(1.661,0.02)(1.651,0.02)(1.632,0.02)(1.622,0.02)(1.607,0.02)(1.610,0.02)(1.556,0.02)(1.541,0.02)(1.532,0.02)(1.523,0.02)(1.417,0.02)(1.426,0.02)(1.436,0.02)(1.443,0.02)(1.509,0.02)(1.458,0.02)(1.498,0.02)(1.469,0.02)(1.482,0.02)(1.487,0.02)(1.401,0.02)(1.392,0.02)(1.378,0.02)(1.366,0.02)(1.271,0.02)(1.291,0.02)(1.296,0.02)(1.343,0.02)(1.340,0.02)(1.336,0.02)(1.324,0.02)(1.318,0.02)(1.307,0.02)(1.259,0.02)(1.255,0.02)(1.245,0.02)(1.240,0.02)(1.230,0.02)(1.223,0.02)(1.209,0.02)(1.202,0.02)(1.193,0.02)(1.089,0.02)(1.178,0.02)(1.170,0.02)(1.166,0.02)(1.103,0.02)(1.110,0.02)(1.149,0.02)(1.136,0.02)(1.126,0.02)(1.119,0.02)(1.158,0.02)(1.078,0.02)(1.070,0.02)(1.062,0.02)(1.057,0.02)(1.042,0.02)(1.034,0.02)(1.027,0.02)(1.020,0.02)(1.011,0.02)(1.000,0.02)(0.995,0.02)(0.987,0.02)(0.980,0.02)(0.972,0.02)(0.956,0.02)(0.947,0.02)(0.951,0.02)(0.862,0.02)(0.869,0.02)(0.934,0.02)(0.928,0.02)(0.918,0.02)(0.880,0.02)(0.908,0.02)(0.891,0.02)(0.897,0.02)(0.902,0.02)(0.854,0.02)(0.672,0.02)(0.674,0.02)(0.841,0.02)(0.844,0.02)(0.837,0.02)(0.829,0.02)(0.818,0.02)(0.815,0.02)(0.688,0.02)(0.691,0.02)(0.806,0.02)(0.798,0.02)(0.790,0.02)(0.784,0.02)(0.775,0.02)(0.704,0.02)(0.767,0.02)(0.712,0.02)(0.714,0.02)(0.723,0.02)(0.759,0.02)(0.733,0.02)(0.737,0.02)(0.748,0.02)(0.745,0.02)(0.629,0.02)(0.638,0.02)(0.646,0.02)(0.664,0.02)(0.657,0.02)(0.660,0.02)(0.625,0.02)(0.619,0.02)(0.612,0.02)(0.604,0.02)(0.535,0.02)(0.595,0.02)(0.546,0.02)(0.563,0.02)(0.556,0.02)(0.571,0.02)(0.578,0.02)(0.552,0.02)(0.584,0.02)(0.587,0.02)(0.588,0.02)(0.542,0.02)(0.528,0.02)(0.522,0.02)(0.513,0.02)(0.508,0.02)(0.504,0.02)(0.500,0.02)(0.495,0.02)(0.489,0.02)(0.481,0.02)(0.478,0.02)(0.388,0.02)(0.393,0.02)(0.474,0.02)(0.464,0.02)(0.397,0.02)(0.404,0.02)(0.407,0.02)(0.470,0.02)(0.456,0.02)(0.451,0.02)(0.415,0.02)(0.419,0.02)(0.437,0.02)(0.430,0.02)(0.445,0.02)(0.448,0.02)(0.427,0.02)(0.425,0.02)(0.262,0.02)(0.265,0.02)(0.381,0.02)(0.377,0.02)(0.369,0.02)(0.366,0.02)(0.273,0.02)(0.276,0.02)(0.353,0.02)(0.360,0.02)(0.349,0.02)(0.345,0.02)(0.361,0.02)(0.282,0.02)(0.284,0.02)(0.287,0.02)(0.291,0.02)(0.295,0.02)(0.302,0.02)(0.306,0.02)(0.310,0.02)(0.319,0.02)(0.338,0.02)(0.332,0.02)(0.330,0.02)(0.324,0.02)(0.326,0.02)(0.249,0.02)(0.252,0.02)(0.256,0.02)(0.209,0.02)(0.243,0.02)(0.241,0.02)(0.236,0.02)(0.234,0.02)(0.214,0.02)(0.218,0.02)(0.221,0.02)(0.226,0.02)(0.223,0.02)(0.228,0.02)(0.205,0.02)(0.201,0.02)(0.197,0.02)(0.191,0.02)(0.188,0.02)(0.196,0.02)(0.148,0.02)(0.150,0.02)(0.182,0.02)(0.178,0.02)(0.180,0.02)(0.155,0.02)(0.158,0.02)(0.171,0.02)(0.162,0.02)(0.167,0.02)(0.165,0.02)(0.173,0.02)(0.087,0.02)(0.090,0.02)(0.093,0.02)(0.092,0.02)(0.098,0.02)(0.100,0.02)(0.102,0.02)(0.108,0.02)(0.113,0.02)(0.109,0.02)(0.118,0.02)(0.119,0.02)(0.124,0.02)(0.127,0.02)(0.129,0.02)(0.137,0.02)(0.131,0.02)(0.133,0.02)(0.141,0.02)(0.144,0.02)(0.143,0.02)(0.160,0.02)(0.052,0.02)(0.056,0.02)(0.059,0.02)(0.062,0.02)(0.060,0.02)(0.068,0.02)(0.071,0.02)(0.064,0.02)(0.085,0.02)(0.081,0.02)(0.077,0.02)(0.074,0.02)(0.076,0.02)(0.083,0.02)(0.109,0.02)(0.066,0.02)(0.053,0.02)(0.029,0.02)(0.048,0.02)(0.030,0.02)(0.046,0.02)(0.045,0.02)(0.042,0.02)(0.041,0.02)(0.040,0.02)(0.037,0.02)(0.033,0.02)(0.034,0.02)(0.035,0.02)(0.026,0.02)(0.016,0.02)(0.019,0.02)(0.017,0.02)(0.022,0.02)(0.023,0.02)(0.024,0.02)
      };
      \legend{{Mar\u{c}enko--Pastur law, $c=5/4$}, {Empirical eigenvalues} }
    \end{axis}
  \end{tikzpicture}
  \caption{Eigenvalues of $\hat{\T}=\T^\oh\X\X^\herm\T^\oh$, where $\T$ is a diagonal of ones but for the first four entries set to $\{3,3,2,2\}$, $N=500$, $n=400$. The theoretical limiting eigenvalue of $\hat{\T}$ is emphasized.}
  \label{fig:spike}
\end{figure}

In Figure~\ref{fig:spike}, we depict the eigenvalues of a spike model as in Theorem~\ref{th:spike1}, with four population eigenvalues greater than one. Two of them exceed the detectability threshold $\omega_i>\sqrt{c}$, while the other two do not. As expected, only two eigenvalues are visible outside the bulk, with values close to their theoretical limit.

Although this was not exactly the approach followed initially in \cite{BAI06}, the same Stieltjes transform and complex integration framework can be used to prove this result. We hereafter give a short description of this proof in the simpler case where $\E=\omega \uu \uu^\herm$ and $\omega>\sqrt{c}$. 

\subsubsection{Extreme eigenvalues}
By definition of an eigenvalue, $\det(\hat{\T}-z\I_N)=0$ for $z$ an eigenvalue of $\hat{\T}$. Some algebraic manipulation based on determinant product formulas and Woodbury's identity \cite{HOR85} lead to
\begin{align*}
	\det(\hat{\T}-z\I_N) &=f_N(z)\det(\I_N+\E)\det(\X\X^\herm-z\I_N) \\
	\text{ with }		  f_N(z) &=1+z\frac{\omega}{1+\omega}\uu^\herm(\X\X^\herm-z\I_N)^{-1}\uu. 
\end{align*}
An eigenvalue of $\hat{\T}$ not inside the main cluster cannot cancel the right-hand determinant in the first line and must therefore cancel $f_N(z)$. Standard random matrix lemmas then show that, since $\uu^\herm(\X\X^\herm-z\I_N)^{-1}\uu$ is asymptotically close to $m_F(z)$, the Stieltjes transform of the Mar\u{c}enko-Pastur law,
\begin{equation*}
	f_N(z) \asto f(z) \triangleq 1+z\frac{\omega}{1+\omega}m_F(z).
\end{equation*}
Substituting $m_F(z)$ by its exact formulation (that is, the Stieltjes transform of the Mar\u{c}enko-Pastur law), we then obtain that $f(z)=0$ is equivalent to $z=1+\omega+c\frac{1+\omega}{\omega}$ if $\omega>\sqrt{c}$ and has no solution otherwise, which is the expected result. 

\subsubsection{Eigenvector projections}
For the result on eigenvector projections, we use again the Cauchy integration scheme which states that, for any deterministic vectors $\aa,\bb\in\CC^N$ and for all large $N$,
\begin{equation}
	\label{eq:complex_int_proj}
	\aa^\herm \hat{\uu}_{1}\hat{\uu}_{1}^\herm \bb = -\frac1{2\pi \imath}\oint_{\mathcal C}\aa^\herm (\hat{\T}-z\I_N)^{-1}\bb
\end{equation}
where $\mathcal C$ is a positively oriented contour circling around $\omega$ and excluding $1$, and where we again recognize a quadratic form involving the resolvent of $\hat{\T}$. This is an immediate consequence of the fact that $(\hat{\T}-z\I_N)^{-1}$ has a pole in $\lambda_1$ with residue $-\hat{\uu}_{1}\hat{\uu}_{1}^\herm$. We then use the same matrix manipulations as in the previous section (by isolating the term $\X\X^\herm$) to finally obtain the result of Theorem~\ref{th:spike1} after residue calculus.

These first order results are important since they suggest detection and identification tests in signal processing scenarios with extremely large arrays. However, for realistic small system sizes, these results are not sufficient to provide efficient statistical tests. To overcome this limitation, we need to go beyond the first order limits and characterize the fluctuations of the extreme eigenvalues and eigenvector projections. Recent works have provided such statistics for the results of Theorem~\ref{th:spike1} as well as for the fluctuations of the estimator \eqref{eq:hatPk}, which we introduce presently. 

\subsection{Fluctuations}
We first mention that the fluctuations of functionals of $F^{\hat{\T}}-F$ in Theorem~\ref{th:BaiSil95} have been derived in \cite{BAI04}. More precisely, for any well-behaved function $f$ (at least holomorphic on the support of $F$) and under some mild technical assumptions,
\begin{equation*}
	N\int f(t)\left(dF^{\hat{\T}}(t)-dF(t)\right)\Rightarrow \mathcal N(0,\sigma^2)
\end{equation*}
with $\sigma^2$ known. This in particular applies to $f(t)=(t-z)^{-1}$ for $z\in\CC\setminus \RR^+$, which gives the asymptotic fluctuations of the Stieltjes transform of $\hat{\T}$. From there, classical asymptotic statistics tools such as the delta-method \cite{VAN00} allow one to transfer the Gaussian fluctuations of the Stieltjes transform $m_{\hat{\T}}$ to the fluctuations of any function of it, e.g. estimators based on $m_{\hat{\T}}$. The following result on the fluctuations of the estimator \eqref{eq:hatPk} therefore comes with no surprise \cite{YAO11}.
\vspace{0.2cm}\begin{theorem}
  \label{th:Jianfeng}
  Consider the estimator $\hat{P}_k$ in \eqref{eq:hatPk}. Assuming the entries of $\X$ have finite fourth order moment (to ensure Theorem~\ref{th:noeig}) and that $P_k$ generates an isolated cluster,
  \begin{equation*}
	  N(\hat{P}_k-P_k) \Rightarrow \mathcal{N}\left(0,\sigma^2\right)
  \end{equation*}
  where $\sigma^2$ is evaluated precisely in \cite{YAO11} as a complex integral form involving derivatives of the limiting Stieltjes transform of $\hat{\T}$.
%  \begin{align*}
%\sigma^2&=-\frac1{4\pi^2 c^2c_k^2} \\ 
%&\oint_{\mathcal C'_k}\oint_{\mathcal C'_k} \left[ \frac{m'(z_1)m'(z_2)}{(m(z_1)-m(z_2))^2}-\frac1{(z_1-z_2)^2}\right]\frac{dz_1 dz_2}{m(z_1)m(z_2)}
%  \end{align*}
\end{theorem}\vspace{0.2cm}

In the case of the spike model, it is of interest to derive the fluctuations of the extreme eigenvalues and eigenvector projections, whose limits were given in Theorem~\ref{th:spike1}. Surprisingly, the fluctuations of these variables are not always Gaussian, \cite{JOH01,BAI05,COU11c}.
\vspace{0.2cm}\begin{theorem}
	\label{th:spike2}
	Let $\hat{\T}$ be defined as in Theorem~\ref{th:spike1}. Then, for $1\leq k\leq K$, if $\omega_k<\sqrt{c}$,
	\begin{align*}
		N^{\frac23}\frac{\lambda_k-(1+\sqrt{c})^2}{(1+\sqrt{c})^{\frac43}\sqrt{c}}\Rightarrow T_2
	\end{align*}
	where $T_2$ is known as the {\it complex Tracy-Widom} distribution, described in \cite{TRA96} as the solution of a Painlev\'e differential equation. If, on the other hand, $\omega_k>\sqrt{c}$, then
\begin{align*}
	\sqrt{N}\begin{pmatrix} |\uu_k^\herm \hat{\uu}_{k}|^2 - \xi_k  \\ \lambda_k - \rho_k \end{pmatrix} \Rightarrow \mathcal{N}\left(0, {\bm\Sigma}_k\right) 
\end{align*} 
where the entries of the $2\times 2$ matrix ${\bf \Sigma}_k$ can be expressed in a simple closed form as a function of $\omega_k$ only.
%\begin{equation*}
%	{\bm\Sigma}_k =
%\begin{bmatrix}
%	\frac{c^2(1+\omega_k)^2}{(c+\omega_k)^2(\omega_k^2-c)}\left(c\frac{(1+\omega_k)^2}{(c+\omega_k)^2} + 1 \right) & \frac{(1+\omega_k)^3c^2}{(\omega_k+c)^2\omega_k} \\
%	\frac{(1+\omega_k)^3c^2}{(\omega_k+c)^2\omega_k} & \frac{c(1+\omega_k)^2(\omega_k^2-c)}{\omega_k^2}	\end{bmatrix}.
%\end{equation*} 
Moreover, for $k\neq k'$ such that $\omega_k,\omega_k'>\sqrt{c}$ and distinct, the centered-scaled eigenvalues and eigenvector projections are asymptotically independent.
\end{theorem}\vspace{0.2cm}

The second result of Theorem~\ref{th:spike2} can be obtained using the same Stieltjes transform and complex integration framework as described in the sketch of proof of Theorem~\ref{th:spike1}, see \cite{COU11c} for details. Again in that case, these are mainly the fluctuations of the Stieltjes transform of $\X\X^\herm$ and an application of the delta-method which lead to the result. The first result of Theorem~\ref{th:spike2} is proved with very different methods which we do not introduce here (since the extreme eigenvalues are inside the base support, complex contour integration approaches cannot be performed). These methods involve techniques such as orthogonal polynomials and Fredholm determinants which rely on the study of the {\it finite} dimensional distribution of the eigenvalues of $\X\X^\herm$ before ultimately providing asymptotic results, see \cite{AND10,FAR06} for details.

\begin{figure}
  \centering
  \begin{tikzpicture}[font=\footnotesize,scale=1]
    \renewcommand{\axisdefaulttryminticks}{4} 
    %\pgfplotsset{every major grid/.append style={densely dashed}}       
    \tikzstyle{every axis y label}+=[yshift=-10pt] 
    \tikzstyle{every axis x label}+=[yshift=5pt]
    \pgfplotsset{every axis legend/.append style={cells={anchor=west},fill=white, at={(0.98,0.98)}, anchor=north east, font=\scriptsize }}
    \begin{axis}[
      %ybar,
      xmin=-5.0,
      ymin=0,
      xmax=5,
      ymax=0.5,
      bar width=4pt,
      grid=major,
      ymajorgrids=false,
      scaled ticks=true,
      %scale ticks above={4},
      xlabel={Centered-scaled largest eigenvalue of $\X\X^\herm$},
      ylabel={Density}
      ]
      \addplot+[ybar,mark=none,color=black,fill=blue!40!white] coordinates{
      (-5.,0.00050)(-4.8,0.00050)(-4.6,0.00250)(-4.4,0.0040)(-4.2,0.0160)(-4.,0.01950)(-3.8,0.03450)(-3.6,0.0620)(-3.4,0.09550)(-3.2,0.15750)(-3.,0.20450)(-2.8,0.30150)(-2.6,0.3270)(-2.4,0.41350)(-2.2,0.41250)(-2.,0.4370)(-1.8,0.430)(-1.6,0.4250)(-1.4,0.3760)(-1.2,0.30)(-1.,0.27150)(-0.8,0.2250)(-0.6,0.1630)(-0.4,0.10050)(-0.2,0.07650)(0.,0.0490)(0.2,0.0360)(0.4,0.02050)(0.6,0.0190)(0.8,0.00850)(1.,0.0030)(1.2,0.00350)(1.4,0.00250)(1.6,0.00150)(1.8,0.00050)(2.,0.0)
      };
      \addplot[smooth,red,line width=0.5pt] plot coordinates{
(-5.0,0.0)(-4.5,0.002)(-4,0.015)
(-3.91,0.022)(-3.87,0.025)(-3.83,0.029)(-3.79,0.033)(-3.75,0.037)(-3.71,0.041)(-3.67,0.046)(-3.63,0.051)(-3.59,0.057)(-3.55,0.064)(-3.51,0.070)(-3.47,0.078)(-3.43,0.085)(-3.39,0.093)(-3.35,0.102)(-3.31,0.111)(-3.27,0.121)(-3.23,0.131)(-3.19,0.142)(-3.15,0.153)(-3.11,0.164)(-3.07,0.176)(-3.03,0.188)(-2.99,0.20)(-2.95,0.213)(-2.91,0.226)(-2.87,0.239)(-2.83,0.252)(-2.79,0.265)(-2.75,0.278)(-2.71,0.291)(-2.67,0.304)(-2.63,0.317)(-2.59,0.329)(-2.55,0.341)(-2.51,0.353)(-2.47,0.364)(-2.43,0.375)(-2.39,0.385)(-2.35,0.395)(-2.31,0.403)(-2.27,0.412)(-2.23,0.419)(-2.19,0.425)(-2.15,0.431)(-2.11,0.436)(-2.07,0.440)(-2.03,0.443)(-1.99,0.446)(-1.95,0.447)(-1.91,0.447)(-1.87,0.447)(-1.83,0.445)(-1.79,0.443)(-1.75,0.440)(-1.71,0.436)(-1.67,0.431)(-1.63,0.426)(-1.59,0.419)(-1.55,0.413)(-1.51,0.405)(-1.47,0.397)(-1.43,0.388)(-1.39,0.379)(-1.35,0.370)(-1.31,0.360)(-1.27,0.350)(-1.23,0.339)(-1.19,0.328)(-1.15,0.317)(-1.11,0.306)(-1.07,0.295)(-1.04,0.286)(-1.02,0.279)(-1.,0.274)(-0.98,0.268)(-0.96,0.263)(-0.94,0.257)(-0.92,0.251)(-0.9,0.246)(-0.88,0.241)(-0.86,0.235)(-0.84,0.229)(-0.82,0.224)(-0.8,0.218)(-0.78,0.214)(-0.76,0.208)(-0.74,0.203)(-0.72,0.198)(-0.7,0.193)(-0.68,0.188)(-0.66,0.183)(-0.64,0.177)(-0.62,0.173)(-0.6,0.167)(-0.58,0.163)(-0.56,0.158)(-0.54,0.154)(-0.52,0.149)(-0.5,0.144)(-0.48,0.140)(-0.46,0.136)(-0.44,0.132)(-0.42,0.128)(-0.4,0.124)(-0.38,0.120)(-0.36,0.117)(-0.34,0.113)(-0.32,0.109)(-0.3,0.105)(-0.28,0.101)(-0.26,0.099)(-0.24,0.095)(-0.22,0.091)(-0.2,0.089)(-0.18,0.086)(-0.16,0.082)(-0.14,0.079)(-0.12,0.077)(-0.1,0.074)(-0.08,0.072)(-0.04,0.067)(-0.02,0.064)(0.,0.061)(0.02,0.059)(0.04,0.056)(0.06,0.055)(0.08,0.052)(0.1,0.050)(0.12,0.048)(0.14,0.047)(0.16,0.044)(0.18,0.043)(0.2,0.042)(0.22,0.039)(0.24,0.038)(0.26,0.037)(0.28,0.035)(0.3,0.034)(0.32,0.032)(0.34,0.031)(0.36,0.029)(0.38,0.028)(0.4,0.027)(0.42,0.026)(0.44,0.024)(0.46,0.023)(0.48,0.022)(0.5,0.022)(0.52,0.020)(0.54,0.020)(0.56,0.018)(0.58,0.017)(0.6,0.017)(0.62,0.017)(0.64,0.016)(0.66,0.015)(0.68,0.014)(0.7,0.013)(0.72,0.013)(0.74,0.012)(0.76,0.012)(0.78,0.011)(0.8,0.010)(0.82,0.010)(0.84,0.009)(0.86,0.009)(0.88,0.009)(0.9,0.008)(0.92,0.008)(0.94,0.007)(0.96,0.006)(0.98,0.007)(1.,0.007)(1.02,0.006)(1.04,0.006)(1.06,0.006)(1.08,0.005)(1.1,0.005)(1.12,0.005)(1.14,0.004)(1.16,0.004)(1.18,0.004)(1.2,0.004)(1.22,0.004)(1.24,0.003)(1.26,0.003)(1.28,0.003)(1.3,0.003)(1.32,0.002)(1.34,0.002)(1.36,0.002)(1.38,0.003)(1.4,0.002)(1.42,0.002)(1.44,0.001)(1.46,0.002)(1.48,0.001)(1.5,0.001)(1.52,0.001)(1.54,0.001)(1.56,0.002)(1.58,0.001)(1.6,0.002)(1.62,0.001)(1.64,0.001)(1.66,0.001)(1.68,0.001)(1.7,0.0)(1.72,0.001)(1.74,0.0)(1.76,0.001)(1.78,0.001)(1.8,0.001)(1.82,0.001)(1.84,0.001)(1.86,0.001)(1.88,0.001)(1.9,0.001)(1.92,0.001)(1.94,0.0)(1.96,0.0)(1.98,0.001)(2.,0.001)(2.04,0.001)(2.06,0.0)
      };
      \legend{{Empirical Eigenvalues},{Tracy-Widom law $T_2$}}
    \end{axis}
  \end{tikzpicture}
  \caption{Distribution of $N^{\frac23}c^{-\frac12}(1+\sqrt{c})^{-\frac43}\left[\lambda_1-(1+\sqrt{c})^2\right]$ against the Tracy-Widom law for $N=500$, $n=1500$, $c=1/3$, for the covariance matrix model $\X\X^\herm$, $\X\in\CC^{N\times n}$ with Gaussian i.i.d. entries. Empirical distribution from $10,000$ Monte-Carlo simulations.}
  \label{fig:tracy-widom}
\end{figure}
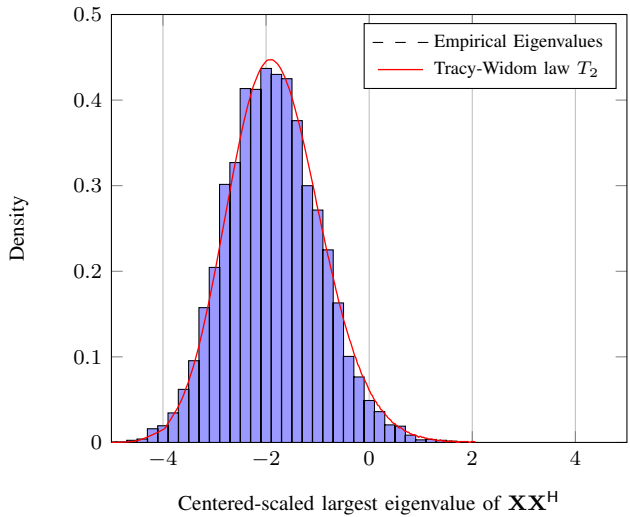

The Tracy-Widom distribution is depicted in Figure~\ref{fig:tracy-widom}. It is interesting to see that the Tracy-Widom law is centered on a negative value and that the probability for positive values is low. This suggests that, if the largest eigenvalue of $\hat{\T}$ is greater than $(1+\sqrt{c})^2$, then it is very likely that $\T$ is not the identity matrix. A similar remark can be made for the smallest eigenvalue of $\hat{\T}$ which has a mirrored Tracy-Widom fluctuation \cite{FEL10}, when $c<1$.

Theorem~\ref{th:spike2} allows for asymptotically accurate test statistics for the decision on signal-plus-noise against pure noise models, thanks to the result on the extreme eigenvalues. Moreover, the results on the eigenvector projections provide even more information to the observer. In the specific context of failure diagnosis \cite{COU11c}, introduced in Section~\ref{sec:applications}, this can be used for the identification of the (asymptotically) most likely assumption from a set of $M$ failure models of the type $(\I_N+\E_k)^\oh\X\X^\herm(\I_N+\E_k)^\oh$, for $k\in\{1,\ldots,M\}$. 

%It is interesting to note that, apart from the proof of the Tracy-Widom fluctuations which requires much different tools, the large dimensional statistics of the extreme eigenvalues given in Theorems~\ref{th:spike1}-\ref{th:spike2} can be all derived using again the Cauchy integration method described in Section~\ref{sec:eigeninference}, see \cite{COU11c}, with integration contours taken around individual extreme eigenvalues. To convince oneself for the simplest case of the limit of the extreme eigenvalues in a spike model, 

\subsection{Practical limitations}
For all techniques derived above, it is important to keep in mind that the results are only valid under the assumption of large matrix dimensions. As it turns out from convergence speed considerations, estimators based on weak convergence properties of the eigenvalue distribution are accurate even for very small system sizes ($N=8$ is often sufficient, if not less), while second order statistics of functionals of the eigenvalue distribution roughly require $N$ to be at least of order $64$. When it comes to asymptotic properties of the extreme eigenvalues, due to the loss of the collective averaging effect of all the eigenvalues, the convergence speed is much slower so that $N=16$ is often a minimum for first order convergence, while $N$ of order $128$ is required for asymptotic second order statistics to become accurate. Note also that these empirical values assume ``non-degenerated'' conditions; for instance, for $N=128$, the second order statistics of the largest eigenvalue in a spike model with population eigenvalue $\omega=\sqrt{c}+\varepsilon$ for a small $\varepsilon>0$ are often far from Gaussian with zero mean (since they should be simultaneously close to follow the very negatively-centered Tracy-Widom distribution) so that, depending on how small $\varepsilon$ is chosen, much larger $N$ are needed for an accurate Gaussian approximation to arise. This suggests that all the above methods have to be manipulated with extreme care depending on the application at hand.

\bigskip

The results introduced so far are only a small subset of the important contributions resulting from more than ten years of applied random matrix theory. A large exposition of sketches of proofs, main strategies to address applied random matrix problems, as well as a large amount of applications are analyzed in the book \cite{COUbook} in the joint context of wireless communications and signal processing. An extension of some notions introduced in the present tutorial can also be found in \cite{COUSP}. Deeper random matrix consideration about the Stieltjes transform approach can be found in \cite{SIL06} and references therein, while discussions on extreme eigenvalues and the tools needed to derive the Tracy-Widom fluctuations, based on the theory of orthogonal polynomials and Fredholm determinants, can be found in e.g. \cite{AND10,FAR06,FYO04} and references therein. In the next section, we discuss several application examples, already partly introduced above, which apply Theorems~\ref{th:BaiSil95}-\ref{th:spike2} in various contexts of hypothesis testing and subspace statistical inference.

\section{Case studies}
\label{sec:applications}

%\begin{figure}
%	\centering
%	\includegraphics[width=8cm]{figs/DoA.pdf}
%	\caption{A radar detection scenario.}
%	\label{fig:DoA}
%\end{figure}

In this section, several classical detection and estimation schemes are revisited for the regime $N\simeq n$ studied in Section~\ref{sec:rmt}.

\subsection{Multi-source power inference in i.i.d. channels}

Our first case study originates from the field of wireless communication, closely related to the previous example of eigenvalue inference. We consider a cognitive sensing scenario \cite{MIT99} in which an $N$-antenna sensing device (a mobile terminal for instance) collects data from $K$ multi-antenna transmitters. Transmitter $k$ is equipped with $M_k$ antennas and sends the signals $\x_{k}(t)\in\CC^{M_k}$ to the receiver at time $t$ through the channel $\H_k\in\CC^{N\times M_k}$, containing i.i.d. Gaussian entries with zero mean and variance $1/N$.\footnote{The normalization of the channel entries by $1/N$ along with $\Vert \x_k(t)\Vert$ bounded allows one to keep the transmit signal power bounded as $N$ grows.} We assume $M_1+\ldots+M_K\triangleq M\leq N$ and that $\H\triangleq [\H_1,\ldots,\H_K]$ remains static during at least $n$ symbol periods. The receiver also captures white Gaussian noise $\sigma\w(t)$ with zero mean and variance $\sigma^2$. The receive signal $\y(t)\in\CC^N$ at time $t$ is then modelled as
\begin{equation*}
	\y(t) = \sum_{i=1}^K \sqrt{P}_i \H_i \x_i(t) + \sigma \w(t).
\end{equation*}
Gathering $n$ realizations of $\y(t)\in\CC^N$ in the matrix $\Y=[\y(1),\ldots,\y(n)]$, $\Y$ takes the form
\begin{equation*}
	\Y = \begin{pmatrix} \H\P^\oh & \sigma\I_N \end{pmatrix}\begin{pmatrix} \X \\ \W \end{pmatrix}
\end{equation*}
where $\X=[\x(1),\ldots,\x(n)]$, $\x(t)=[\x_1(t)^\trans,\ldots,\x_K(t)]^\trans$, $\P=\diag(P_1 \I_{M_1},\ldots,P_K \I_{M_K})$, and $\W=[\w(1),\ldots,\w(n)]$. We wish to estimate the power values $P_1,\ldots,P_K$ from the observation $\Y$. We see that the sample covariance matrix approach developed in Section~\ref{sec:eigeninference} is no longer valid as the population matrix $\T\triangleq \H\P\H^\herm+\sigma^2\I_N$ is now random.

Following the general methodology developed in Section~\ref{sec:eigeninference}, assuming the asymptotic cluster separation property for a certain $P_k$, we first use the connection between $P_k$ and $m_P(z)$, the limiting Stieltjes transform of $\P$, given by \eqref{eq:PkmP}. The innovation from the study in Section~\ref{sec:eigeninference} is that the link between $m_P$ and $m_{\underline F}$, the limiting Stieltjes transform of $\frac1N\Y^\herm\Y$, is more involved and requires some more random matrix arguments than Theorem~\ref{th:BaiSil95} alone. This link, along with a generalization of Theorem~\ref{th:noeig} to account for the randomness in $\T$, induces the following $(N,n)$-consistent estimator $\hat{P}_k$ of $P_k$ \cite{COU11c}
\begin{equation*}
	  \hat{P}_k = \frac{Nn}{M_k(n-N)}\sum_{i\in \mathcal N_k} (\eta_i-\mu_i)
\end{equation*}
where $\mathcal N_k$ is the set of indexes of the eigenvalues of $\frac1n\Y\Y^\herm$ associated to $P_k$, $\eta_1<\ldots<\eta_N$ are the eigenvalues of $\diag(\blambda)-\frac1N\sqrt{\blambda}\sqrt{\blambda}^\trans$ and $\mu_1< \ldots< \mu_N$ are the eigenvalues of $\diag(\blambda)-\frac1n\sqrt{\blambda}\sqrt{\blambda}^\trans$, where $\lambda_1<\ldots<\lambda_N$ are the ordered eigenvalues of $\frac1n\Y\Y^\herm$.

\begin{figure} 
\centering
\begin{tikzpicture}[font=\footnotesize,scale=1]
  \renewcommand{\axisdefaulttryminticks}{4} 
  %\pgfplotsset{every major grid/.append style={densely dashed}}       
  \tikzstyle{every axis y label}+=[yshift=-10pt] 
  \tikzstyle{every axis x label}+=[yshift=5pt]
  \pgfplotsset{every axis legend/.append style={cells={anchor=west},fill=white, at={(0.98,0.98)}, anchor=north east, font=\scriptsize }}
  \begin{axis}[
    %ybar,
    xmin=-5,
    ymin=-20,
    xmax=30,
    ymax=0,
    bar width=3pt,
    grid=major,
    ymajorgrids=false,
    scaled ticks=true,
    %scale ticks above={4},
    xlabel={SNR [dB] ($\sigma^{-2}$)},
    ylabel={Normalized mean square error [dB]}
    ]
    \addplot[smooth,blue,line width=0.5pt] plot coordinates{
    (-5.000000,-2.407977)(-4.000000,-5.758337)(-3.000000,-9.214122)(-2.000000,-12.504550)(-1.000000,-15.188526)(0.000000,-16.780540)(1.000000,-17.586059)(2.000000,-18.016148)(3.000000,-18.228665)(4.000000,-18.495053)(5.000000,-18.615025)(6.000000,-18.770920)(7.000000,-18.851413)(8.000000,-18.904476)(9.000000,-18.985459)(10.000000,-19.005349)(11.000000,-19.068881)(12.000000,-19.120913)(13.000000,-19.065523)(14.000000,-19.099624)(15.000000,-19.112829)(16.000000,-19.126904)(17.000000,-19.113303)(18.000000,-19.176021)(19.000000,-19.161342)(20.000000,-19.151300)(21.000000,-19.162403)(22.000000,-19.173542)(23.000000,-19.175945)(24.000000,-19.200882)(25.000000,-19.140415)(26.000000,-19.218776)(27.000000,-19.197297)(28.000000,-19.201324)(29.000000,-19.202279)(30.000000,-19.173419)
    };
    \addplot[smooth,red,densely dashed,line width=0.5pt] plot coordinates{
    (-5.000000,4.901791)(-4.000000,2.489038)(-3.000000,0.051034)(-2.000000,-2.380407)(-1.000000,-4.752980)(0.000000,-7.006615)(1.000000,-9.050398)(2.000000,-10.833480)(3.000000,-12.356039)(4.000000,-13.528412)(5.000000,-14.473284)(6.000000,-15.217653)(7.000000,-15.730898)(8.000000,-16.169884)(9.000000,-16.471222)(10.000000,-16.713606)(11.000000,-16.945783)(12.000000,-17.071260)(13.000000,-17.210986)(14.000000,-17.312609)(15.000000,-17.347044)(16.000000,-17.401920)(17.000000,-17.433900)(18.000000,-17.455314)(19.000000,-17.470155)(20.000000,-17.503954)(21.000000,-17.522534)(22.000000,-17.527329)(23.000000,-17.566680)(24.000000,-17.596317)(25.000000,-17.550273)(26.000000,-17.546603)(27.000000,-17.623530)(28.000000,-17.550303)(29.000000,-17.585495)(30.000000,-17.604218)
    };
    \legend{{$\hat{P}_3$},{$\hat{P}_3^\infty$}}
  \end{axis}
\end{tikzpicture}
\caption{Normalized mean square error of largest estimated power $P_3$ for $P_1=1/16$, $P_2=1/4$, $P_3=1$, $M_1=M_2=M_3=4$, $M=12$, $N=24$, $n=128$. Comparison between conventional and Stieltjes transform approaches.}
\label{fig:eigen-inferenceMSE_individual2}
\end{figure}
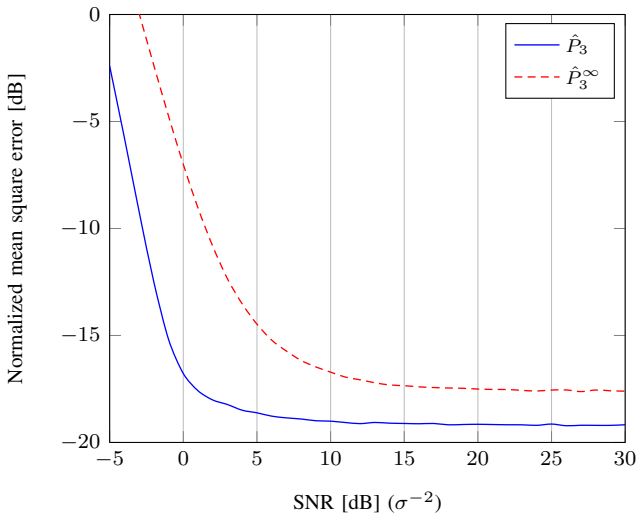

In Figure~\ref{fig:eigen-inferenceMSE_individual2}, the performance comparison between the classical large $n$ approach and the Stieltjes transform approach for power estimation is depicted, based on $10,000$ Monte Carlo simulations. Similar to the estimator \eqref{eq:Pinfty}, the classical approach of $P_k$, denoted $\hat{P}_k^\infty$, consists in assuming $n\gg N$ and $N\gg M_i$ for each $i$ (recall that $M_i$ is the multiplicity of $P_i$). A clear performance gain is observed for the whole range of signal-to-noise ratios (SNR), defined as $\sigma^{-2}$, although a saturation level is observed which translates the approximation error due to the asymptotic analysis. For low SNR, an {\it avalanche} effect is observed, which is caused by the inaccuracy of the Stieltjes transform approach when clusters tend to merge (here the cluster associated to the value $\sigma^2$ grows large as $\sigma^{2}$ increases and covers the clusters associated to $P_1$, then $P_2$, and finally $P_3$). Nonetheless, this apparently strong cluster separability limitation generates an avalanche below the SNR level of the well-known avalanche effect produced by the classical large $n$ estimator. The random matrix method therefore yields a twofold performance gain compared to the classical method, as it is both more accurate (which was expected) and is also more robust to noise which is an interesting, not necessarily anticipated, outcome.

Our next example is based on a contribution from Mestre \cite{MES08c} which initiated a series of statistical inference methods exploiting complex integration, among which the source detection methods described above.

\subsection{G-MUSIC}
The MUSIC method, originally due to Schmidt and later extended by McCloud and Scharf \cite{SCH86,MCC02}, is a direction of arrival detection technique based on a subspace approach. The scenario consists of an array of $N$ sensors (e.g. a radar antenna) receiving the signals originating from $K$ sources at angles $\theta_1,\ldots,\theta_K$ with respect to the array. The objective is to estimate those angles. The data vector $\y(t)$ received at time $t$ by the array can be written as
\begin{equation*}
  %\label{eq:DoA_model}
  \y(t) = \sum_{k=1}^K \s(\theta_k)x_k(t) + \sigma\w(t)
\end{equation*}
where $\s(\theta)\in\CC^N$ is a deterministic vector-valued function of the angle $\theta\in[0,2\pi)$, $x_k(t)\in\CC$ is the proper complex Gaussian i.i.d. data sent by source $k$ at time $t$ and $\w(t)\in\CC^N$ is the additive proper complex Gaussian noise. The vector $\y(t)$ is Gaussian with covariance 
\begin{equation*}
	\T = \S(\Theta)\S(\Theta)^\herm + \sigma^2 \I_N
\end{equation*}
where $\S(\Theta)=[\s(\theta_1),\ldots,\s(\theta_K)]\in\CC^{N\times K}$. We denote as usual $\Y=[\y(1),\ldots,\y(n)]\in\CC^{N\times n}$ the concatenated matrix of $n$ independent observations.

We call $\omega_1\leq \ldots \leq \omega_N$ the eigenvalues of $\T$ and $\uu_1,\ldots,\uu_N$ their associated eigenvectors. Similarly, we will denote $\lambda_1\leq \ldots\leq \lambda_N$ the eigenvalues of $\hat{\T}\triangleq \frac1n\Y\Y^\herm$, with respective eigenvectors $\hat{\uu}_{1},\ldots,\hat{\uu}_{N}$. Assuming $N\geq K$, the smallest $N-K$ eigenvalues of $\T$ equal $\sigma^2$ and we can represent $\T$ under the form
\begin{equation*}
\T = \begin{pmatrix} \U_W & \U_S \end{pmatrix} \begin{pmatrix} \sigma^2\I_{N-K} & {\bf 0} \\ {\bf 0} & {\bm \Omega}_S  \end{pmatrix}\begin{pmatrix} \U_W^\herm \\ \U_S^\herm \end{pmatrix}
\end{equation*}
${\bm \Omega}_S=\diag(\omega_{N-K+1},\ldots,\omega_N)$, $\U_S = [\uu_{N-K+1},\ldots,\uu_N]$ the so-called {\it signal space} and $\U_W = [\uu_1,\ldots,\uu_{N-K}]$ the so-called {\it noise space}.

The basic idea of the MUSIC method is to observe that any vector lying in the signal space is orthogonal to the noise space. This leads in particular to
\begin{equation*}
\eta(\theta_k) \triangleq \s(\theta_k)^\herm\U_W\U_W^\herm \s(\theta_k) = 0
\end{equation*}
for $k\in\{1,\ldots,K\}$.

A natural estimator $\hat{\theta}_k$ of $\theta_k$ in the neighborhood of $\theta_k$ consists in the minimal argument of
\begin{equation*}
	\hat{\eta}(\theta) \triangleq \s(\theta)^\herm\hat\U_{W}\hat\U_{W}^\herm\s(\theta)
\end{equation*}
where $\hat\U_{W}=[\hat{\uu}_{1},\ldots,\hat{\uu}_{N-K}]$ is the eigenvector space corresponding to the smallest $N-K$ eigenvalues of $\hat{\T}$.

This estimator is however proven inconsistent for $N,n$ growing large simultaneously in \cite{MES08}, which is now not surprising. To produce an $(N,n)$-consistent estimator, recall \eqref{eq:complex_int_proj}. Similar to the derivation of the eigenvector projection in Section~\ref{sec:spike}, we can write 
\begin{equation*}
	\s(\theta)^\herm\U_W\U_W^\herm \s(\theta) = \frac1{2\pi \imath}\oint_{\mathcal C_k} \s(\theta)^\herm (\T-z\I_N)^{-1} \s(\theta)dz
\end{equation*}
for a contour $\mathcal C_k$ circling around $\sigma^2$ only. Connecting $\T$ to $\hat{\T}$ from a theorem similar to Theorem~\ref{th:BaiSil95} and performing residue calculus, we finally obtain a good approximation of $\eta(\theta)$, and then an $(N,n)$-consistent estimator for the direction of arrival. This estimator consists precisely in determining the $K$ deepest minima (zeros may not exist) of the function \cite{MES08c}
\begin{equation*}
	\s(\theta)^\herm\left(\sum_{i=1}^N \phi(i) \hat{\uu}_{i}\hat{\uu}_{i}^\herm \right)\s(\theta)
\end{equation*}
with $\phi(i)$ defined as
\begin{align*}
\phi(i) = \left\{ 
\begin{array}{l}
1 + \sum_{k=N-K+1}^N \left[\frac{{\lambda}_k}{{\lambda}_i-{\lambda}_k} - \frac{{\mu}_k}{{\lambda}_i-{\mu}_k} \right],~i\leq N-K \\
- \sum_{k=1}^{N-K} \left[\frac{{\lambda}_k}{{\lambda}_i-{\lambda}_k} - \frac{{\mu}_k}{{\lambda}_i-{\mu}_k} \right],~i> N-K
\end{array}
\right.
\end{align*}
and where, similar to above, $\mu_1\leq \ldots\leq \mu_N$ are the eigenvalues of $\diag({\blambda})-\frac1n\sqrt{{\blambda}}\sqrt{{\blambda}}^\trans$, with ${\blambda}=({\lambda}_1,\ldots,{\lambda}_N)^\trans$, $\lambda_1\leq\ldots\leq\lambda_N$ the ordered eigenvalues of $\hat\T$.

Observe that, contrary to the classical MUSIC method, not only the noise subspace but all the eigenvectors of $\hat{\T}$ are used, although they are divided in two subsets with different weight policies applied to the eigenvectors in each set. This result is generalized to the scenario where $x_k(t)$ is non random. In this case, the technique involves different reference results than the sample covariance matrix theorems required here but the final estimator is obtained similarly \cite{LOU10}. 

\begin{figure}
\centering
\begin{tikzpicture}[font=\footnotesize,scale=1]
\renewcommand{\axisdefaulttryminticks}{8}
%\pgfplotsset{every major grid/.append style={dashed}}
\pgfplotsset{every axis legend/.append style={fill=white,cells={anchor=west},at={(0.02,0.02)},anchor=south west}} \tikzstyle{every axis y label}+=[yshift=-10pt]
\tikzstyle{every axis x label}+=[yshift=5pt]

\begin{axis}[xlabel={Angle [deg]},ylabel={Cost function [dB]},grid=major,ymajorgrids=false,
xmin=33,xmax=39,ymin=-31,ymax=-15,xtick={35,37}
]

\addplot[smooth,red,densely dashed,line width=.5pt] plot coordinates {
(-45.0,2.417819)(-44.640,2.442562)(-44.280,2.516475)(-43.920,2.638610)(-43.560,2.807424)(-43.20,3.020809)(-42.840,3.276157)(-42.480,3.570423)(-42.120,3.900192)(-41.760,4.261760)(-41.40,4.651205)(-41.040,5.064455)(-40.680,5.497357)(-40.320,5.945730)(-39.960,6.405414)(-39.60,6.872312)(-39.240,7.342421)(-38.880,7.811853)(-38.520,8.276856)(-38.160,8.733825)(-37.80,9.179312)(-37.440,9.610030)(-37.080,10.022864)(-36.720,10.414876)(-36.360,10.783315)(-36.0,11.125637)(-35.640,11.439527)(-35.280,11.722932)(-34.920,11.974110)(-34.560,12.191694)(-34.20,12.374772)(-33.840,12.522992)(-33.480,12.636682)(-33.120,12.716998)(-32.760,12.766067)(-32.40,12.787126)(-32.040,12.784614)(-31.680,12.764165)(-31.320,12.732445)(-30.960,12.696781)(-30.60,12.664536)(-30.240,12.642282)(-29.880,12.634891)(-29.520,12.644754)(-29.160,12.671369)(-28.80,12.711439)(-28.440,12.759521)(-28.080,12.809050)(-27.720,12.853534)(-27.360,12.887655)(-27.0,12.908133)(-26.640,12.914220)(-26.280,12.907778)(-25.920,12.892896)(-25.560,12.875062)(-25.20,12.859969)(-24.840,12.852185)(-24.480,12.854044)(-24.120,12.865101)(-23.760,12.882395)(-23.40,12.901451)(-23.040,12.917707)(-22.680,12.927912)(-22.320,12.931074)(-21.960,12.928656)(-21.60,12.923943)(-21.240,12.920727)(-20.880,12.921713)(-20.520,12.927283)(-20.160,12.935174)(-19.80,12.941339)(-19.440,12.941738)(-19.080,12.934385)(-18.720,12.920795)(-18.360,12.906094)(-18.0,12.897370)(-17.640,12.900568)(-17.280,12.917058)(-16.920,12.941638)(-16.560,12.963373)(-16.20,12.969396)(-15.840,12.950442)(-15.480,12.906207)(-15.120,12.848448)(-14.760,12.799547)(-14.40,12.784549)(-14.040,12.817444)(-13.680,12.887693)(-13.320,12.955312)(-12.960,12.956926)(-12.60,12.816236)(-12.240,12.450023)(-11.880,11.764443)(-11.520,10.637790)(-11.160,8.877318)(-10.80,6.097452)(-10.440,1.228617)(-10.080,-12.142106)(-9.720,-2.524199)(-9.360,4.365792)(-9.0,7.879591)(-8.640,10.040901)(-8.280,11.423839)(-7.920,12.281175)(-7.560,12.758463)(-7.20,12.961459)(-6.840,12.982949)(-6.480,12.911196)(-6.120,12.824624)(-5.760,12.777346)(-5.40,12.787256)(-5.040,12.838987)(-4.680,12.901335)(-4.320,12.946799)(-3.960,12.962902)(-3.60,12.953511)(-3.240,12.933142)(-2.880,12.918092)(-2.520,12.918405)(-2.160,12.934173)(-1.80,12.957409)(-1.440,12.977512)(-1.080,12.986750)(-0.720,12.983183)(-0.360,12.970380)(0.0,12.954845)(0.360,12.942736)(0.720,12.937440)(1.080,12.938933)(1.440,12.944818)(1.80,12.952128)(2.160,12.958823)(2.520,12.964367)(2.880,12.969361)(3.240,12.974674)(3.60,12.980607)(3.960,12.986521)(4.320,12.991030)(4.680,12.992644)(5.040,12.990548)(5.40,12.985160)(5.760,12.978230)(6.120,12.972343)(6.480,12.969931)(6.840,12.972155)(7.20,12.978144)(7.560,12.985041)(7.920,12.988983)(8.280,12.986695)(8.640,12.977065)(9.0,12.961985)(9.360,12.945994)(9.720,12.934663)(10.080,12.932278)(10.440,12.939815)(10.80,12.954260)(11.160,12.969704)(11.520,12.979767)(11.880,12.980258)(12.240,12.970957)(12.60,12.955765)(12.960,12.941148)(13.320,12.933385)(13.680,12.935770)(14.040,12.947042)(14.40,12.961854)(14.760,12.973052)(15.120,12.974686)(15.480,12.964498)(15.840,12.944932)(16.20,12.922339)(16.560,12.904603)(16.920,12.898023)(17.280,12.904684)(17.640,12.921511)(18.0,12.941430)(18.360,12.956079)(18.720,12.958895)(19.080,12.947448)(19.440,12.924330)(19.80,12.896384)(20.160,12.872459)(20.520,12.860298)(20.880,12.863592)(21.240,12.880351)(21.60,12.903269)(21.960,12.921897)(22.320,12.925722)(22.680,12.907132)(23.040,12.863561)(23.40,12.798512)(23.760,12.721344)(24.120,12.645711)(24.480,12.586661)(24.840,12.55680)(25.20,12.562529)(25.560,12.601729)(25.920,12.663817)(26.280,12.731995)(26.640,12.786544)(27.0,12.807850)(27.360,12.778401)(27.720,12.683651)(28.080,12.512051)(28.440,12.254607)(28.80,11.904247)(29.160,11.455175)(29.520,10.902265)(29.880,10.240522)(30.240,9.464592)(30.60,8.568276)(30.960,7.544025)(31.320,6.382339)(31.680,5.071016)(32.040,3.594153)(32.40,1.930734)(32.760,0.052634)(33.120,-2.078203)(33.480,-4.513387)(33.840,-7.320741)(34.20,-10.574720)(34.560,-14.274959)(34.920,-17.955357)(35.280,-20.079096)(35.640,-20.120574)(36.0,-19.687729)(36.360,-19.517199)(36.720,-19.307923)(37.080,-18.336728)(37.440,-16.374112)(37.80,-13.971977)(38.160,-11.636501)(38.520,-9.552457)(38.880,-7.742714)(39.240,-6.181029)(39.60,-4.832556)(39.960,-3.665357)(40.320,-2.652757)(40.680,-1.773093)(41.040,-1.008888)(41.40,-0.346024)(41.760,0.226944)(42.120,0.719327)(42.480,1.138708)(42.840,1.491252)(43.20,1.781939)(43.560,2.014744)(43.920,2.192765)(44.280,2.318321)(44.640,2.393022)(45.0,2.417819)
};
\addplot[blue,smooth,line width=.5pt] plot coordinates {
(-45.0,2.229542)(-44.640,2.255122)(-44.280,2.331517)(-43.920,2.457698)(-43.560,2.631988)(-43.20,2.852113)(-42.840,3.115265)(-42.480,3.418184)(-42.120,3.757239)(-41.760,4.128512)(-41.40,4.527884)(-41.040,4.951106)(-40.680,5.393871)(-40.320,5.851870)(-39.960,6.320843)(-39.60,6.796615)(-39.240,7.275125)(-38.880,7.752449)(-38.520,8.224813)(-38.160,8.688606)(-37.80,9.140379)(-37.440,9.576857)(-37.080,9.994937)(-36.720,10.391699)(-36.360,10.764414)(-36.0,11.110556)(-35.640,11.427830)(-35.280,11.714202)(-34.920,11.967948)(-34.560,12.187716)(-34.20,12.372609)(-33.840,12.522286)(-33.480,12.637089)(-33.120,12.718188)(-32.760,12.767726)(-32.40,12.788966)(-32.040,12.786377)(-31.680,12.765637)(-31.320,12.733474)(-30.960,12.697286)(-30.60,12.664518)(-30.240,12.641827)(-29.880,12.634158)(-29.520,12.643948)(-29.160,12.670707)(-28.80,12.711115)(-28.440,12.759672)(-28.080,12.809740)(-27.720,12.854748)(-27.360,12.889307)(-27.0,12.910085)(-26.640,12.916308)(-26.280,12.909839)(-25.920,12.894799)(-25.560,12.876729)(-25.20,12.861389)(-24.840,12.853417)(-24.480,12.855197)(-24.120,12.866304)(-23.760,12.883764)(-23.40,12.903056)(-23.040,12.919560)(-22.680,12.929963)(-22.320,12.933226)(-21.960,12.930793)(-21.60,12.925959)(-21.240,12.922552)(-20.880,12.923332)(-20.520,12.928733)(-20.160,12.936527)(-19.80,12.942680)(-19.440,12.943137)(-19.080,12.935879)(-18.720,12.922390)(-18.360,12.907770)(-18.0,12.899102)(-17.640,12.902343)(-17.280,12.918881)(-16.920,12.943530)(-16.560,12.965356)(-16.20,12.971474)(-15.840,12.952591)(-15.480,12.908376)(-15.120,12.850573)(-14.760,12.801576)(-14.40,12.786465)(-14.040,12.819277)(-13.680,12.889505)(-13.320,12.957157)(-12.960,12.958809)(-12.60,12.818089)(-12.240,12.451683)(-11.880,11.765604)(-11.520,10.637859)(-11.160,8.874933)(-10.80,6.088435)(-10.440,1.190568)(-10.080,-13.180632)(-9.720,-2.621909)(-9.360,4.349632)(-9.0,7.874859)(-8.640,10.039710)(-8.280,11.424130)(-7.920,12.282198)(-7.560,12.759895)(-7.20,12.963149)(-6.840,12.984820)(-6.480,12.913191)(-6.120,12.826678)(-5.760,12.779387)(-5.40,12.789221)(-5.040,12.840849)(-4.680,12.903111)(-4.320,12.948531)(-3.960,12.964634)(-3.60,12.955267)(-3.240,12.934913)(-2.880,12.919851)(-2.520,12.920129)(-2.160,12.935871)(-1.80,12.959129)(-1.440,12.979320)(-1.080,12.988698)(-0.720,12.985272)(-0.360,12.972547)(0.0,12.956984)(0.360,12.944739)(0.720,12.939250)(1.080,12.940575)(1.440,12.946388)(1.80,12.953753)(2.160,12.960604)(2.520,12.966336)(2.880,12.971474)(3.240,12.976836)(3.60,12.982724)(3.960,12.988533)(4.320,12.992934)(4.680,12.994482)(5.040,12.992373)(5.40,12.987011)(5.760,12.980115)(6.120,12.974246)(6.480,12.971833)(6.840,12.974047)(7.20,12.980029)(7.560,12.986924)(7.920,12.990849)(8.280,12.988502)(8.640,12.978757)(9.0,12.963521)(9.360,12.947382)(9.720,12.935979)(10.080,12.933652)(10.440,12.941392)(10.80,12.956139)(11.160,12.971897)(11.520,12.982191)(11.880,12.982767)(12.240,12.973389)(12.60,12.958008)(12.960,12.943175)(13.320,12.935260)(13.680,12.937607)(14.040,12.948948)(14.40,12.963878)(14.760,12.975154)(15.120,12.976760)(15.480,12.966414)(15.840,12.946595)(16.20,12.923734)(16.560,12.905811)(16.920,12.899196)(17.280,12.905988)(17.640,12.923060)(18.0,12.943241)(18.360,12.958065)(18.720,12.960897)(19.080,12.949289)(19.440,12.925877)(19.80,12.897601)(20.160,12.873424)(20.520,12.861183)(20.880,12.864611)(21.240,12.881680)(21.60,12.904987)(21.960,12.923951)(22.320,12.927941)(22.680,12.909262)(23.040,12.865328)(23.40,12.799685)(23.760,12.721802)(24.120,12.645486)(24.480,12.585954)(24.840,12.555945)(25.20,12.561913)(25.560,12.601684)(25.920,12.664525)(26.280,12.733440)(26.640,12.788508)(27.0,12.809947)(27.360,12.780111)(27.720,12.684351)(28.080,12.511032)(28.440,12.251069)(28.80,11.897284)(29.160,11.443738)(29.520,10.885115)(29.880,10.216156)(30.240,9.431140)(30.60,8.523353)(30.960,7.484508)(31.320,6.304031)(31.680,4.968110)(32.040,3.458342)(32.40,1.749666)(32.760,-0.192945)(33.120,-2.420143)(33.480,-5.008523)(33.840,-8.081042)(34.20,-11.850902)(34.560,-16.716622)(34.920,-23.239410)(35.280,-27.411952)(35.640,-24.626772)(36.0,-23.254516)(36.360,-23.879766)(36.720,-26.134105)(37.080,-26.526848)(37.440,-21.721926)(37.80,-17.132539)(38.160,-13.648526)(38.520,-10.939720)(38.880,-8.762384)(39.240,-6.968405)(39.60,-5.464369)(39.960,-4.188092)(40.320,-3.096184)(40.680,-2.157186)(41.040,-1.347588)(41.40,-0.64940)(41.760,-0.048599)(42.120,0.465888)(42.480,0.902876)(42.840,1.269413)(43.20,1.571113)(43.560,1.812413)(43.920,1.996741)(44.280,2.126650)(44.640,2.203904)(45.0,2.229542)
};
\legend{{MUSIC},{G-MUSIC}}
\end{axis}
\end{tikzpicture}
\caption{MUSIC against G-MUSIC for DoA detection of $K=2$ signal sources, $N=20$ sensors, $n=150$ samples, SNR of $10$ dB. Angles of arrival of $35^\circ$ and $37^\circ$.}
\label{fig:applications_estimation_DoA_zoom}
\end{figure}
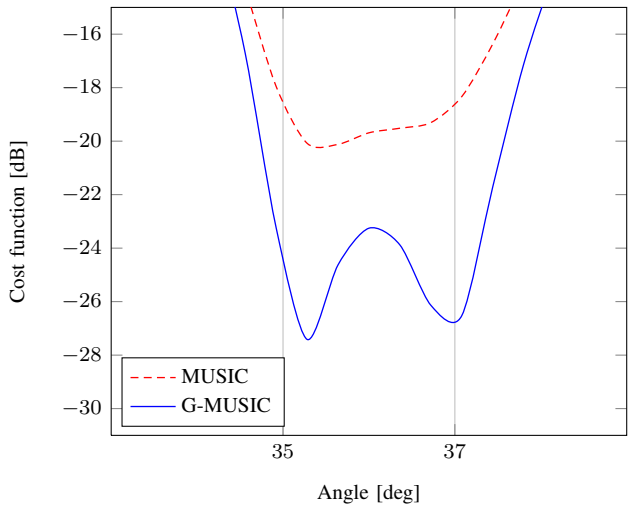

In Figure~\ref{fig:applications_estimation_DoA_zoom}, the performance comparison between the traditional MUSIC and the G-MUSIC algorithms is depicted, when two sources have close angles which the traditional MUSIC algorithm is not able to distinguish. It is clear on this single-shot realization that the G-MUSIC algorithm shows deeper minima and allows for a better source resolution.

Note that both inference methods described above, be it the power or angle of arrival estimation schemes, assume a priori knowledge of the cluster indexes to be able to implement the estimators. The proposed random matrix method therefore does not solve the {\it order selection} problem. New techniques are therefore being investigated that deal with the generalization of the Akaike principle \cite{AKA74} and the minimum description length technique \cite{RIS83} for improved order selection using random matrix theory, see e.g. \cite{NAD10}. 

Remark also that, so far, only regular sample covariance matrix models have been analyzed to address questions of array processing. In reality, due (in particular) to the non-Gaussian noise structure in radar detection, more advanced estimators of the population covariance matrix are used based on the celebrated Huber and Maronna robust M-estimators \cite{MAR76,kent1991redescending}. We are currently investigating these M-estimators within the random matrix framework. Further intricate models involving multi-path propagation and the technique consisting in stacking successive vector observations are also under study.

This completes the set of examples using results based on weak limits of large dimensional random matrices and G-estimation. We now turn to examples involving the results for the spike models, starting with simple signal detection procedures. 

\subsection{Detection}
\label{sec:detection}
Consider the hypothesis test which consists in deciding whether a received signal $\y(t)\in\CC^N$ consists of pure noise, $\y(t)=\sigma \w(t)$, with $\w(t)\in\CC^N$ with Gaussian entries with zero mean and unit variance, for some unknown $\sigma$ (hypothesis $\mathcal H_0$), or containing a signal plus noise, $\y(t)=\h x(t)+\sigma \w(t)$, for a time-independent vector channel $\h\in\CC^N$, a scalar proper complex Gaussian signal $x(t)$ (hypothesis $\mathcal H_1$) and a noise variance $\sigma^2$, which are all unknown. As usual, we gather $n$ i.i.d. received data in $\Y=[\y(1),\ldots,\y(n)]$.

Since $\h$ is unknown, we cannot compare directly $\mathcal H_0$ and $\mathcal H_1$. Instead, we will accept or reject $\mathcal H_0$ based on how $\Y$ fits the pure noise hypothesis. It is a known result that the generalized likelihood ratio test (GLRT) for the decision between $\mathcal H_0$ and $\mathcal H_1$ boils down to the test \cite{WAX85}
\begin{equation*}
	\lambda'_1 \triangleq \frac{\lambda_1}{(\frac1{Nn}\tr\Y\Y^\herm)} \overset{ {\mathcal H}_0}{\underset{{\mathcal H}_1}{\lessgtr}} f(\varepsilon)
\end{equation*}
with $\lambda_1$ the largest eigenvalue of $\frac1n\Y\Y^\herm$, for $f$ a given monotonic function and $\varepsilon$ the maximally acceptable false alarm rate. Evaluating the statistical properties of $\lambda'_1$ for finite $N$ is however rather involved and leads to impractical tests, see e.g. \cite{RAT05}. Instead, we consider here a very elementary statistical test based on Theorem~\ref{th:spike2}.

It is clear that $\frac1{Nn}\tr\Y\Y^\herm\asto \sigma^2$ under both $\mathcal H_0$ or $\mathcal H_1$. Therefore, an application of Slutsky's lemma \cite{VAN00} ensures that the asymptotic fluctuations of $\lambda'_1$ follow a Tracy-Widom distribution around $(1+\sqrt{N/n})^2$. The GLRT method therefore leads asymptotically to test the Tracy-Widom statistics for the appropriately centered and scaled version of $\lambda'_1$. More precisely, for a false alarm rate $\varepsilon$, this is
\begin{equation*}
	N^{\frac23}\frac{\lambda'_1-(1+\sqrt{c})^2}{(1+\sqrt{c})^{\frac43}\sqrt{c}} \overset{ {\mathcal H}_0}{\underset{{\mathcal H}_1}{\lessgtr}} T_2^{-1}\left( 1 - \varepsilon \right)
\end{equation*}
with $c=N/n$ and $T_2$ the Tracy-Widom distribution. Further properties of the above statistical test, and in particular theoretical expression of false alarm rates and test powers are derived in \cite{BIA10} using the theory of large deviations. Due to the relatively slow convergence speed of the largest eigenvalue distribution to the Tracy-Widom law, some recent works have proposed refinements, see e.g. \cite{NAD11}.
 
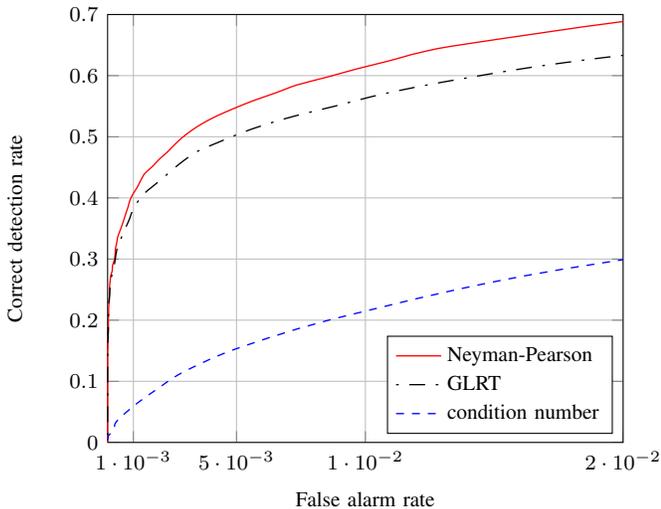
\begin{figure}
\centering
\begin{tikzpicture}[font=\footnotesize,scale=1]
\renewcommand{\axisdefaulttryminticks}{8}
%\pgfplotsset{every major grid/.append style={dashed}}
\pgfplotsset{every axis legend/.append style={fill=white,cells={anchor=west},at={(0.98,0.02)},anchor=south east}} \tikzstyle{every axis y label}+=[yshift=-10pt]
\tikzstyle{every axis x label}+=[yshift=5pt]
\tikzstyle{dashed dotted}=[dash pattern=on 1pt off 4pt on 6pt off 4pt]

\begin{axis}[xlabel=False alarm rate,ylabel=Correct detection rate
,grid=major,
xmin=0,xmax=0.02,ymin=0.0,ymax=0.7,xtick={0.001,0.005,0.01,0.02}
]

\addplot[smooth,red,line width=.5pt] plot coordinates {
(0.024550,0.69) (0.020980,0.693) 
(0.013780,0.650930) (0.012080,0.637360) (0.010890,0.623860) (0.009650,0.610840) (0.008540,0.597870) (0.007340,0.584670) (0.006520,0.571830) (0.005630,0.558690) (0.004910,0.546450) (0.004240,0.534120) (0.003690,0.522020) (0.003250,0.510140) (0.002880,0.498150) (0.002580,0.486390) (0.002320,0.475110) (0.002010,0.463550) (0.001740,0.451430) (0.001440,0.439910) (0.001290,0.429080) (0.001160,0.418160) (0.001000,0.407420) (0.000870,0.396610) (0.000800,0.386060) (0.000720,0.376170) (0.000640,0.365610) (0.000560,0.355510) (0.000470,0.345260) (0.000390,0.335650) (0.000350,0.325590) (0.000300,0.316200) (0.000290,0.306970) (0.000260,0.298120) (0.000190,0.289410) (0.000180,0.280850) (0.000120,0.272650) (0.000100,0.264430) (0.000080,0.256600) (0.000070,0.248460) (0.000070,0.240830) (0.000070,0.233650) (0.000050,0.226290) (0.000040,0.219200) (0.000040,0.211920) (0.000030,0.204820) (0.000030,0.197950) (0.000030,0.191180) (0.000030,0.184840) (0.000030,0.178760) (0.000010,0.172750) (0.000010,0.166740) (0.000010,0.161140) (0.000010,0.155650) (0,0.150370) (0,0.145030) (0,0.139940) (0,0.134990) (0,0.130000) (0,0.125630) (0,0.121280) (0,0.117070) (0,0.113280) (0,0.108990) (0,0.105260) (0,0.101190) (0,0.097210) (0,0.093780) (0,0.090480) (0,0.086870) (0,0.083490) (0,0.080240) (0,0.077500) (0,0.074440) (0,0.071660) (0,0.068990) (0,0.066310) (0,0.063460) (0,0.060910) (0,0.058280) (0,0.055780) (0,0.053530) (0,0.051220) (0,0.049060) (0,0.046800) (0,0.044730) (0,0.042660) (0,0.040720) (0,0.039150) (0,0.037650) (0,0.035980) (0,0.034530) (0,0.032880) (0,0.031690) (0,0.030060) (0,0.028710) (0,0.027410) (0,0.026070) (0,0.024920) (0,0.023710) (0,0.022740) (0,0.021780) (0,0.020800) (0,0.019800) (0,0.018750) (0,0.017930) (0,0.017130) (0,0.016420) (0,0.015720) (0,0.015030) (0,0.014310) (0,0.013640) (0,0.013050) (0,0.012300) (0,0.011730) (0,0.011030) (0,0.010390) (0,0.009880) (0,0.009360) (0,0.008930) (0,0.008560) (0,0.008100) (0,0.007690) (0,0.007250) (0,0.006900) (0,0.006490) (0,0.006190) (0,0.005830) (0,0.005560) (0,0.005170) (0,0.004900) (0,0.004580) (0,0.004320) (0,0.004100) (0,0.003760) (0,0.003520) (0,0.003280) (0,0.003140) (0,0.002980) (0,0.002760) (0,0.002600) (0,0.002440) (0,0.002300) (0,0.002200) (0,0.002070) (0,0.001930) (0,0.001820) (0,0.001700) (0,0.001610) (0,0.001520) (0,0.001330) (0,0.001240) (0,0.001140) (0,0.001060) (0,0.000960) (0,0.000920) (0,0.000880) (0,0.000810) (0,0.000730) (0,0.000670) (0,0.000640) (0,0.000610) (0,0.000550) (0,0.000510) (0,0.000490) (0,0.000460) (0,0.000430) (0,0.000400) (0,0.000400) (0,0.000380) (0,0.000340) (0,0.000300) (0,0.000270) (0,0.000250) (0,0.000200) (0,0.000190) (0,0.000180) (0,0.000180) (0,0.000160) (0,0.000160) (0,0.000140) (0,0.000130) (0,0.000100) (0,0.000100) (0,0.000090) (0,0.000080) (0,0.000080) (0,0.000070) (0,0.000060) (0,0.000060) (0,0.000050) (0,0.000040) (0,0.000040) (0,0.000020) (0,0.000020) (0,0.000020) (0,0.000020) (0,0.000020) (0,0.000020) (0,0.000020) (0,0.000010) (0,0) 
};

\addplot[smooth,black,dashed dotted,line width=.5pt] plot coordinates {
(0.999990,1.000000) (0.999840,1.000000) (0.999010,0.999940) (0.995530,0.999760) (0.986900,0.999270) (0.970660,0.998300) (0.945370,0.996530) (0.911750,0.994340) (0.868510,0.990770) (0.819520,0.986690) (0.763490,0.981090) (0.705600,0.974540) (0.646570,0.967040) (0.587180,0.958550) (0.529600,0.949080) (0.475600,0.938770) (0.423830,0.927670) (0.375280,0.916620) (0.330170,0.904620) (0.290120,0.891510) (0.253340,0.878360) (0.221150,0.864500) (0.191750,0.850420) (0.165190,0.835410) (0.142500,0.820480) (0.122210,0.805430) (0.104580,0.790180) (0.089480,0.774870) (0.076700,0.59050) (0.065540,0.743320) (0.055920,0.727440) (0.047330,0.711870) (0.040360,0.695740) (0.033710,0.679490) (0.027850,0.663290) (0.023100,0.646770) (0.019370,0.630110) (0.016490,0.613980) (0.014100,0.597470) (0.012000,0.581580) (0.010260,0.565850) (0.008830,0.550210) (0.007380,0.535300) (0.006120,0.520040) (0.005120,0.505220) (0.004270,0.490480) (0.003480,0.476080) (0.002960,0.461250) (0.002520,0.446430) (0.002150,0.432550) (0.001740,0.418560) (0.001370,0.405260) (0.001140,0.391970) (0.000960,0.378640) (0.000850,0.365600) (0.000710,0.352290) (0.000580,0.340200) (0.000470,0.327510) (0.000360,0.315460) (0.000310,0.303870) (0.000270,0.292820) (0.000210,0.281690) (0.000150,0.270910) (0.000090,0.260810) (0.000090,0.251030) (0.000080,0.241110) (0.000050,0.231780) (0.000040,0.222280) (0.000040,0.213120) (0.000030,0.203840) (0.000030,0.195100) (0.000030,0.187120) (0.000020,0.179150) (0.000020,0.171360) (0.000010,0.164330) (0,0.157110) (0,0.150150) (0,0.143560) (0,0.136880) (0,0.130620) (0,0.124950) (0,0.119170) (0,0.114100) (0,0.109300) (0,0.104040) (0,0.099230) (0,0.094310) (0,0.089750) (0,0.085820) (0,0.081780) (0,0.077860) (0,0.074220) (0,0.070450) (0,0.066760) (0,0.063310) (0,0.059920) (0,0.056670) (0,0.053510) (0,0.050830) (0,0.048080) (0,0.045390) (0,0.043160) (0,0.040940) (0,0.038600) (0,0.036210) (0,0.034190) (0,0.032210) (0,0.030480) (0,0.028860) (0,0.027230) (0,0.025920) (0,0.024420) (0,0.022950) (0,0.021340) (0,0.020080) (0,0.018840) (0,0.017820) (0,0.016790) (0,0.015910) (0,0.014990) (0,0.014150) (0,0.013350) (0,0.012400) (0,0.011730) (0,0.010930) (0,0.010320) (0,0.009710) (0,0.009130) (0,0.008520) (0,0.007990) (0,0.007430) (0,0.006870) (0,0.006300) (0,0.005860) (0,0.005480) (0,0.005080) (0,0.004770) (0,0.004320) (0,0.004040) (0,0.003820) (0,0.003590) (0,0.003360) (0,0.003060) (0,0.002780) (0,0.002590) (0,0.002440) (0,0.002250) (0,0.002070) (0,0.001930) (0,0.001780) (0,0.001570) (0,0.001400) (0,0.001290) (0,0.001200) (0,0.001090) (0,0.001010) (0,0.000910) (0,0.000840) (0,0.000790) (0,0.000740) (0,0.000670) (0,0.000610) (0,0.000540) (0,0.000490) (0,0.000450) (0,0.000410) (0,0.000350) (0,0.000310) (0,0.000290) (0,0.000230) (0,0.000180) (0,0.000170) (0,0.000170) (0,0.000170) (0,0.000140) (0,0.000130) (0,0.000120) (0,0.000100) (0,0.000090) (0,0.000080) (0,0.000080) (0,0.000040) (0,0.000020) (0,0.000020) (0,0.000020) (0,0.000020) (0,0.000020) (0,0.000010) (0,0.000010) (0,0.000010) (0,0.000010) (0,0.000010) (0,0.000010) (0,0.000010) (0,0.000010) (0,0.000010) (0,0.000010) (0,0.000010) (0,0.000010) (0,0)
};

\addplot[smooth,blue,dashed,line width=.5pt] plot coordinates {
(0.999990,1.000000) (0.672840,0.955570) (0.248160,0.770960) (0.092870,0.566240) (0.039540,0.404260) (0.018470,0.288860) (0.009330,0.207980) (0.004970,0.152910) (0.002880,0.113810) (0.001870,0.085860) (0.001190,0.065510) (0.000750,0.050360) (0.000450,0.039010) (0.000290,0.030810) (0.000260,0.024550) (0.000190,0.019740) (0.000140,0.016260) (0.000070,0.012890) (0.000030,0.010630) (0.000020,0.008900) (0.000020,0.007250) (0.000010,0.006030) (0,0.005140) (0,0.004410) (0,0.003790) (0,0.003350) (0,0.002810) (0,0.002240) (0,0.001920) (0,0.001710) (0,0.001530) (0,0.001410) (0,0.001200) (0,0.001070) (0,0.000990) (0,0.000890) (0,0.000780) (0,0.000690) (0,0.000640) (0,0.000580) (0,0.000510) (0,0.000450) (0,0.000440) (0,0.000380) (0,0.000360) (0,0.000330) (0,0.000310) (0,0.000300) (0,0.000270) (0,0.000250) (0,0.000230) (0,0.000220) (0,0.000190) (0,0.000170) (0,0.000140) (0,0.000120) (0,0.000120) (0,0.000120) (0,0.000110) (0,0.000100) (0,0.000090) (0,0.000080) (0,0.000080) (0,0.000070) (0,0.000070) (0,0.000070) (0,0.000070) (0,0.000070) (0,0.000070) (0,0.000070) (0,0.000070) (0,0.000070) (0,0.000070) (0,0.000070) (0,0.000070) (0,0.000070) (0,0.000060) (0,0.000050) (0,0.000050) (0,0.000050) (0,0.000050) (0,0.000050) (0,0.000050) (0,0.000050) (0,0.000040) (0,0.000040) (0,0.000030) (0,0.000030) (0,0.000030) (0,0.000020) (0,0.000020) (0,0.000020) (0,0.000020) (0,0.000020) (0,0.000020) (0,0.000020) (0,0.000020) (0,0.000020) (0,0.000020) (0,0.000020) (0,0.000020) (0,0.000020) (0,0.000020) (0,0.000020) (0,0.000020) (0,0.000020) (0,0.000020) (0,0.000020) (0,0.000020) (0,0.000020) (0,0.000020) (0,0.000020) (0,0.000020) (0,0.000020) (0,0.000020) (0,0.000020) (0,0.000020) (0,0.000020) (0,0.000020) (0,0.000020) (0,0.000020) (0,0.000020) (0,0.000020) (0,0.000020) (0,0.000020) (0,0.000020) (0,0.000020) (0,0.000020) (0,0.000020) (0,0.000020) (0,0.000020) (0,0.000020) (0,0.000020) (0,0.000020) (0,0.000020) (0,0.000020) (0,0.000020) (0,0.000020) (0,0.000020) (0,0.000020) (0,0.000010) (0,0.000010) (0,0.000010) (0,0.000010) (0,0.000010) (0,0.000010) (0,0.000010) (0,0.000010) (0,0.000010) (0,0.000010) (0,0.000010) (0,0.000010) (0,0.000010) (0,0.000010) (0,0.000010) (0,0.000010) (0,0.000010) (0,0.000010) (0,0.000010) (0,0.000010) (0,0.000010) (0,0.000010) (0,0.000010) (0,0.000010) (0,0.000010) (0,0.000010) (0,0.000010) (0,0.000010) (0,0.000010) (0,0.000010) (0,0.000010) (0,0.000010) (0,0.000010) (0,0.000010) (0,0.000010) (0,0.000010) (0,0.000010) (0,0.000010) (0,0.000010) (0,0.000010) (0,0.000010) (0,0.000010) (0,0.000010) (0,0.000010) (0,0.000010) (0,0.000010) (0,0.000010) (0,0.000010) (0,0.000010) (0,0.000010) (0,0.000010) (0,0.000010) (0,0.000010) (0,0.000010) (0,0.000010) (0,0.000010) (0,0.000010) (0,0.000010) (0,0.000010) (0,0)
};

\legend{Neyman-Pearson \\ GLRT \\ condition number \\};
\end{axis}
\end{tikzpicture}
\caption{Receiver operating curve for a priori unknown $\sigma^2$ of the Neyman-Pearson test (N-P), condition number method and GLRT, $N=4$, $n=8$, SNR$=0~{\rm dB}$, $\h$ has Gaussian entries of zero mean and unit variance. For the Neyman-Pearson test, both uniform and Jeffreys prior, with exponent $\beta=1$, are provided.}
\label{fig:unknownSNR}
\end{figure}

In Figure~\ref{fig:unknownSNR}, the performance of the GLRT detector given by the receiver operating curve for different false alarm rate levels is depicted, for small system sizes. We present a comparison between the GLRT approach, the empirical technique \cite{CAR08} which consists in using the conditioning number of $\frac1n\Y\Y^\herm$ (ratio largest and the smallest eigenvalues) as a signal detector and the optimal Neyman-Pearson detector (with knowledge of the channel statistics) derived in \cite{COU09b}. It turns out that the suboptimal asymptotic GLRT approach is quite close in performance to the finite dimensional exact optimum, the latter being however quite complex to implement.

Our last application uses the second order statistics of eigenvalues and eigenvectors for a spike model in the context of failure localization in large dimensional networks.

\subsection{Failure detection in large networks}

The method presented here allows for fast detection and localization of local failures (few links or few nodes) in a large sensor network of $N$ sensors gathering data about $M$ system parameters $\theta_1,\ldots,\theta_M$. Assume the linear scenario 
\begin{equation*}
	\x(t) = \H {\bm\theta}(t) + \sigma \w(t) 
\end{equation*}
for ${\bm\theta}(t)=[\theta_1(t),\ldots,\theta_M(t)]^\trans\in\CC^M$ and $\w(t)\in\CC^N$ two complex Gaussian signals with independent entries of zero mean and unit variance. The objective here is to detect a small rank perturbation of $\EE[\x(t)\x(t)^\herm]\triangleq \T=\H\H^\herm + \sigma^2\I_N$, originating from a local failure or change in the parameters, assuming $\T$ known. In particular, when the entry $k$ of ${\bm\theta}(t)$ shows a sudden change in variance, the model changes as
\begin{equation*}
	\x'(t) = \H(\I_M+\alpha_k \e_k\e_k^\herm) {\bm\theta}(t) + \sigma \w(t) 
\end{equation*}
for a given $\alpha_k\geq -1$ and with $\e_k\in\CC^M$ the vector of all zeros but for $\e_k(k)=1$. Taking for instance $\alpha_k=-1$ turns the entry $k$ of ${\bm\theta}(t)$ into zero, corresponding to a complete collapse of the parameter under control. Denoting $\y(t)=\T^{-\oh}\x(t)$ and $\y'(t)=\T^{-\oh}\x'(t)$, we have $\EE[\y(t)\y(t)^\herm]=\I_N$, while
\begin{equation*}
	\EE[\y'(t)\y^{\prime\herm}(t)] = \I_N + [(1+\alpha_k)^2-1] \T^{-\oh}\H\e_k\e_k^\herm \H^\herm \T^{-\oh}
\end{equation*}
which is a rank-$1$ perturbation of $\I_N$.

Now assume that we do not know whether the model follows the expression of $\y(t)$ or $\y'(t)$, and let us generically denote both by $\y(t)$. 
A simple off-line failure (or change) detection test consists, as in Section~\ref{sec:detection}, to decide whether the largest eigenvalue of $\frac1n\Y\Y^\herm$, with $\Y=[\y(1),\ldots,\y(n)]$, has a Tracy-Widom distribution. However, we wish now to go further and to be able to decide, upon failure detection, which entry $\theta_k$ of ${\bm\theta}$ was altered. For this, we need to provide a statistical test for the hypothesis $\mathcal H_k=$``parameter $\theta_k$ failed''. A mere maximum likelihood procedure consisting in testing the Gaussian distribution of $\Y$, assuming $\alpha_k$ known for each $k$, is however costly for $N$ large and becomes impractical if $\alpha_k$ is unknown. Instead, one can perform a statistical test on the extreme eigenvalues and eigenvector projections of $\frac1n\Y\Y^\herm$, which mainly costs the computational time of eigenvalue decomposition (already performed in the detection test). For this, we need to assume that the number $n$ of observations is sufficiently large for a failure of any parameter of ${\bm\theta}$ to be detectable. As an immediate application of Theorem~\ref{th:spike2}, denoting $\lambda$ the largest eigenvalue of $\frac1n\Y\Y^\herm$ and $\hat\uu$ its corresponding eigenvector, the estimator $\hat{k}$ for $k$ is then given by \cite{COU11c}
\begin{align*}
	\hat{k} = \arg\max_{1\leq i\leq M} &-N\begin{pmatrix} |\uu_i^\herm \hat\uu|^2 - \xi_i  \\ \lambda - \rho_i \end{pmatrix}^\trans {\bm\Sigma}_i^{-1} \begin{pmatrix} |\uu_i^\herm \hat\uu|^2 - \xi_i  \\ \lambda - \rho_i \end{pmatrix} \\ 
		&- \log\det {\bm\Sigma}_i
\end{align*}
with $\uu_i$ such that $\Vert\uu_i\Vert=1$, $\omega_i\uu_i\uu_i^\herm=[(1+\alpha_i)^2-1] \T^{-\oh}\H\e_i\e_i^\herm \H^\herm \T^{-\oh}$, and $\xi_i$, $\rho_i$, ${\bm\Sigma}_i$ defined as a function of $\omega_i$ in Theorem~\ref{th:spike2} (the index $i$ refers here to a change in parameter $\theta_i$ and not to the $i$-th largest eigenvalue of the small rank perturbation matrix).

This procedure still assumes that $\alpha_i$ is known for each $i$. If not, an estimator of $\alpha_i$ can be derived and Theorem~\ref{th:spike2} adapted accordingly to account for the fluctuations of the estimator.

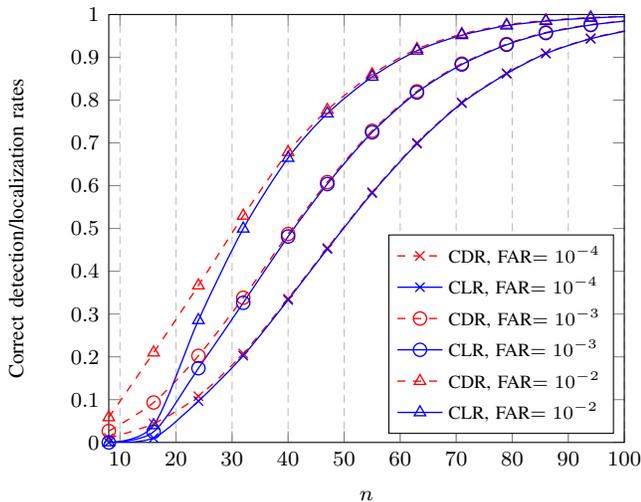
\begin{figure}
  \centering
  \begin{tikzpicture}[font=\footnotesize,scale=1]
    \renewcommand{\axisdefaulttryminticks}{8} 
    \pgfplotsset{every axis/.append style={mark options=solid, mark size=2.5pt}}
    \tikzstyle{every major grid}+=[style=densely dashed]       
%    \tikzstyle{every pin}=[fill=white,draw=black,font=\footnotesize,edge style={<-}]     
%\pgfplotsset{every axis y label/.append style={yshift=-20pt}} 
%\pgfplotsset{every axis x label/.append style={yshift=5pt}} 
    %\tikzstyle{every axis x label}+=[yshift=5pt]
    \tikzstyle{every axis legend}+=[cells={anchor=west},fill=white,%draw=none,
        at={(0.98,0.02)}, anchor=south east, font=\scriptsize ]

    \begin{axis}[
      xmajorgrids=true,
      xlabel={$n$},
      ylabel={Correct detection/localization rates},
      xmin=8,
      xmax=100, 
      ymin=0, 
      ymax=1,
      ]
      \addplot[smooth,dashed,mark=x,red,line width=0.5pt] plot coordinates{
      (8,0.010200)(16,0.043800)(24,0.107960)(32,0.207860)(40,0.336030)(47,0.454400)(55,0.584990)(63,0.700300)(71,0.794630)(79,0.862870)(86,0.909710)(94,0.944260)(102,0.966340)(110,0.981400)(118,0.989790)(125,0.994050)(133,0.996800)(141,0.998470)(149,0.999110)(157,0.999680)(164,0.999820)
      };
      \addplot[smooth,blue,mark=x,line width=0.5pt] plot coordinates{
      (8,0.000000)(16,0.010520)(24,0.096780)(32,0.202810)(40,0.333010)(47,0.452000)(55,0.582830)(63,0.698180)(71,0.792920)(79,0.861600)(86,0.908710)(94,0.943300)(102,0.965540)(110,0.980880)(118,0.989350)(125,0.993700)(133,0.996450)(141,0.998210)(149,0.998970)(157,0.999550)(164,0.999620)
      };
      \addplot[smooth,dashed,mark=o,red,line width=0.5pt] plot coordinates{
      (8,0.026660)(16,0.093370)(24,0.202310)(32,0.338200)(40,0.487050)(47,0.608330)(55,0.728300)(63,0.820520)(71,0.885220)(79,0.930940)(86,0.958050)(94,0.976350)(102,0.987440)(110,0.993380)(118,0.996930)(125,0.998020)(133,0.999030)(141,0.999610)(149,0.999840)(157,0.999940)(164,0.999940)
      };
      \addplot[smooth,mark=*,blue,mark=o,line width=0.5pt] plot coordinates{
      (8,0.000000)(16,0.024870)(24,0.173410)(32,0.326340)(40,0.481210)(47,0.603840)(55,0.724740)(63,0.817780)(71,0.883170)(79,0.929270)(86,0.956680)(94,0.975450)(102,0.986740)(110,0.992800)(118,0.996450)(125,0.997750)(133,0.998750)(141,0.999310)(149,0.999710)(157,0.999800)(164,0.999730)
      };
      \addplot[smooth,dashed,mark=triangle,red,line width=0.5pt] plot coordinates{
      (8,0.058580)(16,0.209900)(24,0.366650)(32,0.529530)(40,0.678360)(47,0.776730)(55,0.860030)(63,0.919240)(71,0.954200)(79,0.975700)(86,0.986000)(94,0.992830)(102,0.996510)(110,0.998370)(118,0.999240)(125,0.999570)(133,0.999880)(141,0.999920)(149,0.999980)(157,0.999970)(164,1.000000)
      };
      \addplot[smooth,mark=*,blue,mark=triangle,line width=0.5pt] plot coordinates{
      (8,0.000000)(16,0.039970)(24,0.285590)(32,0.499420)(40,0.664690)(47,0.768560)(55,0.854330)(63,0.915330)(71,0.951510)(79,0.973860)(86,0.984460)(94,0.991870)(102,0.995710)(110,0.997770)(118,0.998890)(125,0.999170)(133,0.999620)(141,0.999650)(149,0.999820)(157,0.999830)(164,0.999870)
      };
      \legend{{CDR, FAR$=10^{-4}$},{CLR, FAR$=10^{-4}$},{CDR, FAR$=10^{-3}$},{CLR, FAR$=10^{-3}$},{CDR, FAR$=10^{-2}$},{CLR, FAR$=10^{-2}$}}
    \end{axis}
  \end{tikzpicture}
  \caption{Correct detection (CDR) and localization (CLR) rates for different levels of false alarm rates (FAR) and different values of $n$, for drop of variance of $\theta_1$. The minimal theoretical $n$ for observability is $n=8$.}
  \label{fig:simu10}
\end{figure}

In Figure~\ref{fig:simu10}, we depict the detection and localization performance for a sudden drop to zero of the parameter $\theta_1(t)$ in a scenario with $M=N=10$. We see in Figure~\ref{fig:simu10} that, compared to the detection performance, the localization performance is rather poor for small $n$ but reaches the same level as the detection performance for larger $n$, which is explained by the inaccuracy of the Gaussian approximation for $\omega_i$ close to $\sqrt{c}$. 

\section{Conclusions}
\label{sec:conclusion}
In this short tutorial about statistical inference using large dimensional random matrix theory, we have argued that many traditional signal processing methods are inconsistent when both the population and system dimensions are large. We then introduced notions of random matrix theory which provide results on the spectrum of large random matrices. These results were then used to adjust some of these inconsistent signal processing methods to new consistent estimates. To this end, we presented a recent method based (i) on the Stieltjes transform to derive weak convergence properties of the spectrum of large matrices and (ii) on complex integration to derive estimators. This somewhat parallels the Fourier transform and M-estimator framework usually met in classical asymptotic signal processing \cite{VAN00}. 

Nonetheless, while classical signal processing tools are very mature, statistical inference based on random matrix analysis still lacks many fundamental mathematical results that only experts can provide to this day for advanced system models. The main consequence is the slow appearance of methods to derive various estimators for more involved signal processing problems. The emergence of the aforementioned Stieltjes transform and complex integration framework is a rare exception to this rule which, we believe, is the foundation for many future breakthroughs in this area. 

\appendices

\bibliography{../tutorial_RMT/book_final/IEEEabrv,../tutorial_RMT/book_final/IEEEconf,../tutorial_RMT/book_final/tutorial_RMT,robust_est.bib}

\begin{thebibliography}{10}
\providecommand{\url}[1]{#1}
\csname url@samestyle\endcsname
\providecommand{\newblock}{\relax}
\providecommand{\bibinfo}[2]{#2}
\providecommand{\BIBentrySTDinterwordspacing}{\spaceskip=0pt\relax}
\providecommand{\BIBentryALTinterwordstretchfactor}{4}
\providecommand{\BIBentryALTinterwordspacing}{\spaceskip=\fontdimen2\font plus
\BIBentryALTinterwordstretchfactor\fontdimen3\font minus
  \fontdimen4\font\relax}
\providecommand{\BIBforeignlanguage}[2]{{%
\expandafter\ifx\csname l@#1\endcsname\relax
\typeout{** WARNING: IEEEtran.bst: No hyphenation pattern has been}%
\typeout{** loaded for the language `#1'. Using the pattern for}%
\typeout{** the default language instead.}%
\else
\language=\csname l@#1\endcsname
\fi
#2}}
\providecommand{\BIBdecl}{\relax}
\BIBdecl

\bibitem{VAN00}
A.~W. {Van~der~Vaart}, \emph{{Asymptotic Statistics}}.\hskip 1em plus 0.5em
  minus 0.4em\relax New York: Cambridge University Press, 2000.

\bibitem{MES08c}
X.~Mestre and M.~Lagunas, ``{Modified subspace algorithms for DoA estimation
  with large arrays},'' \emph{{IEEE} Transactions on Signal Processing},
  vol.~56, no.~2, pp. 598--614, Feb. 2008.

\bibitem{POT00}
L.~Laloux, P.~Cizeau, M.~Potters, and J.~P. Bouchaud, ``{Random matrix theory
  and financial correlations},'' \emph{International Journal of Theoretical and
  Applied Finance}, vol.~3, no.~3, pp. 391--397, Jul. 2000.

\bibitem{HAN96}
N.~Hansen and A.~Ostermeier, ``{Adapting arbitrary normal mutation
  distributions in evolution strategies: the covariance matrix adaptation},''
  in \emph{Evolutionary Computation, Proceedings of IEEE International
  Conference on}.\hskip 1em plus 0.5em minus 0.4em\relax IEEE, 1996, pp.
  312--317.

\bibitem{WIG55}
E.~Wigner, ``{Characteristic vectors of bordered matrices with infinite
  dimensions},'' \emph{The Annals of Mathematics}, vol.~62, no.~3, pp.
  548--564, Nov. 1955.

\bibitem{MAR67}
V.~A. Mar\u{c}enko and L.~A. Pastur, ``{Distributions of eigenvalues for some
  sets of random matrices},'' \emph{Math USSR-Sbornik}, vol.~1, no.~4, pp.
  457--483, Apr. 1967.

\bibitem{RUD86}
W.~Rudin, \emph{{Real and Complex Analysis}}, 3rd~ed.\hskip 1em plus 0.5em
  minus 0.4em\relax McGraw-Hill Series in Higher Mathematics, May 1986.

\bibitem{GIR90}
V.~L. Girko, \emph{{Theory of Random Determinants}}.\hskip 1em plus 0.5em minus
  0.4em\relax Dordrecht, The Netherlands: Kluwer, Kluwer Academic Publishers,
  1990.

\bibitem{BIC08}
P.~J. Bickel and E.~Levina, ``Regularized estimation of large covariance
  matrices,'' \emph{The Annals of Statistics}, vol.~36, no.~1, pp. 199--227,
  2008.

\bibitem{WU09}
W.~B. Wu and M.~Pourahmadi, ``Banding sample autocovariance matrices of
  stationary processes,'' \emph{Statistica Sinica}, vol.~19, no.~4, pp.
  1755--1768, 2009.

\bibitem{WIS28}
J.~Wishart, ``{The generalized product moment distribution in samples from a
  normal multivariate population},'' \emph{Biometrika}, vol.~20, no. 1-2, pp.
  32--52, Dec. 1928.

\bibitem{JAM64}
A.~T. James, ``{Distributions of matrix variates and latent roots derived from
  normal samples},'' \emph{The Annals of Mathematical Statistics}, vol.~35,
  no.~2, pp. 475--501, 1964.

\bibitem{RAT05}
T.~Ratnarajah, R.~Vaillancourt, and M.~Alvo, ``{Eigenvalues and condition
  numbers of complex random matrices},'' \emph{SIAM Journal on Matrix Analysis
  and Applications}, vol.~26, no.~2, pp. 441--456, 2005.

\bibitem{PET06}
F.~Hiai and D.~Petz, \emph{{The Semicircle Law, Free Random Variables and
  Entropy - Mathematical Surveys and Monographs No. 77}}.\hskip 1em plus 0.5em
  minus 0.4em\relax Providence, RI, USA: American Mathematical Society, 2006.

\bibitem{BIA03}
P.~Biane, ``{Free probability for probabilists},'' \emph{Quantum Probability
  Communications}, vol.~11, pp. 55--71, 2003.

\bibitem{RYA09b}
A.~Masucci, {\O}.~Ryan, S.~Yang, and M.~Debbah, ``{Finite dimensional
  statistical inference},'' \emph{{IEEE} Transactions on Information Theory},
  vol.~57, no.~4, pp. 2457--2473, 2011.

\bibitem{RAO07}
N.~R. Rao and A.~Edelman, ``{The polynomial method for random matrices},''
  \emph{Foundations of Computational Mathematics}, vol.~8, no.~6, pp. 649--702,
  Dec. 2008.

\bibitem{PAS00b}
L.~A. Pastur and V.~Vasilchuk, ``{On the law of addition of random matrices},''
  \emph{Communications in Mathematical Physics}, vol. 214, no.~2, pp. 249--286,
  2000.

\bibitem{HAC06}
W.~Hachem, O.~Khorunzhy, P.~Loubaton, J.~Najim, and L.~A. Pastur, ``{A new
  approach for capacity analysis of large dimensional multi-antenna
  channels},'' \emph{{IEEE} Transactions on Information Theory}, vol.~54,
  no.~9, pp. 3987--4004, 2008.

\bibitem{SIL95}
J.~W. Silverstein and Z.~D. Bai, ``{On the empirical distribution of
  eigenvalues of a class of large dimensional random matrices},'' \emph{Journal
  of Multivariate Analysis}, vol.~54, no.~2, pp. 175--192, 1995.

\bibitem{MES08b}
X.~Mestre, ``{Improved estimation of eigenvalues of covariance matrices and
  their associated subspaces using their sample estimates},'' \emph{{IEEE}
  Transactions on Information Theory}, vol.~54, no.~11, pp. 5113--5129, Nov.
  2008.

\bibitem{HOR85}
R.~A. Horn and C.~R. Johnson, \emph{{Matrix Analysis}}.\hskip 1em plus 0.5em
  minus 0.4em\relax Cambridge University Press, 1985.

\bibitem{SIL06}
Z.~D. Bai and J.~W. Silverstein, \emph{{Spectral analysis of large dimensional
  random matrices}}, 2nd~ed.\hskip 1em plus 0.5em minus 0.4em\relax New York,
  NY, USA: Springer Series in Statistics, 2009.

\bibitem{AND10}
G.~W. Anderson, A.~Guionnet, and O.~Zeitouni, \emph{{An introduction to random
  matrices}}.\hskip 1em plus 0.5em minus 0.4em\relax Cambridge University
  Press, 2010.

\bibitem{COUbook}
R.~Couillet and M.~Debbah, \emph{{Random Matrix Methods for Wireless
  Communications}}, 1st~ed.\hskip 1em plus 0.5em minus 0.4em\relax New York,
  NY, USA: Cambridge University Press, 2011.

\bibitem{YAT95}
R.~D. Yates, ``{A framework for uplink power control in cellular radio
  systems},'' \emph{{IEEE} Journal on Selected Areas in Communications},
  vol.~13, no.~7, pp. 1341--1347, 1995.

\bibitem{DUP09}
\BIBentryALTinterwordspacing
F.~Dupuy and P.~Loubaton, ``{Mutual information of frequency selective MIMO
  systems: an asymptotic approach},'' 2009. [Online]. Available:
  \url{http://www-syscom.univ-mlv.fr/\~fdupuy/publications.php}
\BIBentrySTDinterwordspacing

\bibitem{COU09}
R.~Couillet, M.~Debbah, and J.~W. Silverstein, ``{A deterministic equivalent
  for the analysis of correlated MIMO multiple access channels},'' \emph{{IEEE}
  Transactions on Information Theory}, vol.~57, no.~6, pp. 3493--3514, Jun.
  2011.

\bibitem{HAC07}
W.~Hachem, P.~Loubaton, and J.~Najim, ``{Deterministic equivalents for certain
  functionals of large random matrices},'' \emph{Annals of Applied
  Probability}, vol.~17, no.~3, pp. 875--930, 2007.

\bibitem{COU11}
\BIBentryALTinterwordspacing
R.~Couillet, J.~Hoydis, and M.~Debbah, ``{Deterministic equivalents for the
  analysis of unitary precoded systems},'' \emph{{IEEE} Transactions on
  Information Theory}, 2011, submitted for publication. [Online]. Available:
  \url{http://arxiv.org/abs/1011.3717}
\BIBentrySTDinterwordspacing

\bibitem{COU11d}
J.~Hoydis, R.~Couillet, and M.~Debbah, ``{Random beamforming over correlated
  fading channels},'' \emph{{IEEE} Transactions on Information Theory}, 2011,
  submitted for publication.

\bibitem{DUP10}
\BIBentryALTinterwordspacing
F.~Dupuy and P.~Loubaton, ``{On the capacity achieving covariance matrix for
  frequency selective MIMO channels using the asymptotic approach},''
  \emph{{IEEE} Transactions on Information Theory}, 2010, to appear. [Online].
  Available: \url{http://arxiv.org/abs/1001.3102}
\BIBentrySTDinterwordspacing

\bibitem{TSE99}
D.~N.~C. Tse and S.~V. Hanly, ``{Linear multiuser receivers: effective
  interference, effective bandwidth and user capacity},'' \emph{{IEEE}
  Transactions on Information Theory}, vol.~45, no.~2, pp. 641--657, Feb. 1999.

\bibitem{SHA99}
S.~{Verd\'u} and S.~Shamai, ``{Spectral efficiency of CDMA with random
  spreading},'' \emph{{IEEE} Transactions on Information Theory}, vol.~45,
  no.~2, pp. 622--640, Feb. 1999.

\bibitem{GIR00}
\BIBentryALTinterwordspacing
V.~L. Girko, ``{Ten years of general statistical analysis}.'' [Online].
  Available:
  \url{www.general-statistical-analysis.girko.freewebspace.com/chapter14.pdf}
\BIBentrySTDinterwordspacing

\bibitem{YIN88b}
J.~W. Silverstein, Z.~D. Bai, and Y.~Q. Yin, ``{A note on the largest
  eigenvalue of a large dimensional sample covariance matrix},'' \emph{Journal
  of Multivariate Analysis}, vol.~26, no.~2, pp. 166--168, 1988.

\bibitem{YIN88}
Y.~Q. Yin, Z.~D. Bai, and P.~R. Krishnaiah, ``{On the limit of the largest
  eigenvalue of the large dimensional sample covariance matrix},''
  \emph{Probability Theory and Related Fields}, vol.~78, no.~4, pp. 509--521,
  1988.

\bibitem{SIL98}
Z.~D. Bai and J.~W. Silverstein, ``{No eigenvalues outside the support of the
  limiting spectral distribution of large dimensional sample covariance
  matrices},'' \emph{The Annals of Probability}, vol.~26, no.~1, pp. 316--345,
  Jan. 1998.

\bibitem{BAI99}
------, ``{Exact separation of eigenvalues of large dimensional sample
  covariance matrices},'' \emph{The Annals of Probability}, vol.~27, no.~3, pp.
  1536--1555, 1999.

\bibitem{DOZ07}
B.~Dozier and J.~W. Silverstein, ``{On the empirical distribution of
  eigenvalues of large dimensional information plus noise-type matrices},''
  \emph{Journal of Multivariate Analysis}, vol.~98, no.~4, pp. 678--694, 2007.

\bibitem{BEN09}
F.~Benaych-Georges and R.~Rao, ``{The eigenvalues and eigenvectors of finite,
  low rank perturbations of large random matrices},'' \emph{Advances in
  Mathematics}, vol. 227, no.~1, pp. 494--521, 2011.

\bibitem{BAI08b}
\BIBentryALTinterwordspacing
Z.~D. Bai and J.~F. Yao, ``{Limit theorems for sample eigenvalues in a
  generalized spiked population model},'' 2008. [Online]. Available:
  \url{http://arxiv.org/abs/0806.1141}
\BIBentrySTDinterwordspacing

\bibitem{BAI06}
J.~Baik and J.~W. Silverstein, ``{Eigenvalues of large sample covariance
  matrices of spiked population models},'' \emph{Journal of Multivariate
  Analysis}, vol.~97, no.~6, pp. 1382--1408, 2006.

\bibitem{COU11c}
R.~Couillet and W.~Hachem, ``{Local failure detection and diagnosis in large
  sensor networks},'' \emph{{IEEE} Transactions on Information Theory}, 2011,
  submitted for publication.

\bibitem{BAI04}
Z.~D. Bai and J.~W. Silverstein, ``{CLT of linear spectral statistics of large
  dimensional sample covariance matrices},'' \emph{The Annals of Probability},
  vol.~32, no.~1A, pp. 553--605, 2004.

\bibitem{YAO11}
J.~Yao, R.~Couillet, J.~Najim, and M.~Debbah, ``{Fluctuations of an Improved
  Population Eigenvalue Estimator in Sample Covariance Matrix Models},''
  \emph{{IEEE} Transactions on Information Theory}, 2011, submitted for
  publication.

\bibitem{JOH01}
I.~M. Johnstone, ``{On the distribution of the largest eigenvalue in principal
  components analysis},'' \emph{Annals of Statistics}, vol.~99, no.~2, pp.
  295--327, 2001.

\bibitem{BAI05}
J.~Baik, G.~{Ben Arous}, and S.~{P\'ech\'e}, ``{Phase transition of the largest
  eigenvalue for non-null complex sample covariance matrices},'' \emph{The
  Annals of Probability}, vol.~33, no.~5, pp. 1643--1697, 2005.

\bibitem{TRA96}
C.~A. Tracy and H.~Widom, ``{On orthogonal and symplectic matrix ensembles},''
  \emph{Communications in Mathematical Physics}, vol. 177, no.~3, pp. 727--754,
  1996.

\bibitem{FAR06}
\BIBentryALTinterwordspacing
J.~Faraut, ``{Random matrices and orthogonal polynomials},'' Lecture Notes,
  CIMPA School of Merida, 2006. [Online]. Available:
  \url{www.math.jussieu.fr/\~faraut/Merida.Notes.pdf}
\BIBentrySTDinterwordspacing

\bibitem{FEL10}
O.~N. Feldheim and S.~Sodin, ``{A universality result for the smallest
  eigenvalues of certain sample covariance matrices},'' \emph{Geometric And
  Functional Analysis}, vol.~20, no.~1, pp. 88--123, 2010.

\bibitem{COUSP}
T.~Chen, D.~Rajan, and E.~Serpedin, \emph{{Mathematical foundations for signal
  processing, communications and networking}}.\hskip 1em plus 0.5em minus
  0.4em\relax Cambridge University Press, 2011.

\bibitem{FYO04}
Y.~V. Fyodorov, ``{Introduction to the random matrix theory: Gaussian unitary
  ensemble and beyond},'' \emph{Recent Perspectives in Random Matrix Theory and
  Number Theory}, vol. 322, pp. 31--78, 2005.

\bibitem{MIT99}
J.~{Mitola~III} and G.~Q. {Maguire~Jr}, ``{Cognitive radio: making software
  radios more personal},'' \emph{{IEEE} Personal Communication Magazine},
  vol.~6, no.~4, pp. 13--18, 1999.

\bibitem{SCH86}
R.~Schmidt, ``{Multiple emitter location and signal parameter estimation},''
  \emph{IEEE Transactions on Antennas and Propagation}, vol.~34, no.~3, pp.
  276--280, 1986.

\bibitem{MCC02}
M.~L. McCloud and L.~L. Scharf, ``{A new subspace identification algorithm for
  high-resolution DOA estimation},'' \emph{IEEE Transactions on Antennas and
  Propagation}, vol.~50, no.~10, pp. 1382--1390, 2002.

\bibitem{MES08}
X.~Mestre, ``{On the asymptotic behavior of the sample estimates of eigenvalues
  and eigenvectors of covariance matrices},'' \emph{{IEEE} Transactions on
  Signal Processing}, vol.~56, no.~11, pp. 5353--5368, Nov. 2008.

\bibitem{LOU10}
\BIBentryALTinterwordspacing
P.~Vallet, P.~Loubaton, and X.~Mestre, ``{Improved subspace estimation for
  multivariate observations of high dimension: the deterministic signals
  case},'' \emph{{IEEE} Transactions on Information Theory}, 2010, submitted
  for publication. [Online]. Available: \url{http://arxiv.org/abs/1002.3234}
\BIBentrySTDinterwordspacing

\bibitem{AKA74}
H.~Akaike, ``A new look at the statistical model identification,'' \emph{{IEEE}
  Transactions Autom. Control}, vol.~19, no.~6, pp. 716--723, 1974.

\bibitem{RIS83}
J.~Rissanen, ``A universal prior for integers and estimation by minimum
  description length,'' \emph{The Annals of Statistics}, vol.~11, no.~2, pp.
  416--431, 1983.

\bibitem{NAD10}
B.~Nadler, ``Nonparametric detection of signals by information theoretic
  criteria: performance analysis and an improved estimator,'' \emph{{IEEE}
  Transactions on Signal Processing}, vol.~58, no.~5, pp. 2746--2756, 2010.

\bibitem{MAR76}
R.~A. Maronna, ``{Robust M-estimators of multivariate location and scatter},''
  \emph{The annals of statistics}, pp. 51--67, 1976.

\bibitem{kent1991redescending}
J.~T. Kent and D.~E. Tyler, ``{Redescending M-estimates of multivariate
  location and scatter},'' \emph{The Annals of Statistics}, pp. 2102--2119,
  1991.

\bibitem{WAX85}
M.~Wax and T.~Kailath, ``{Detection of signals by information theoretic
  criteria},'' \emph{IEEE Transactions on Signal, Speech and Signal
  Processing}, vol.~33, no.~2, pp. 387--392, 1985.

\bibitem{BIA10}
P.~Bianchi, J.~Najim, M.~Maida, and M.~Debbah, ``{Performance of some
  eigen-based hypothesis tests for collaborative sensing},'' \emph{{IEEE}
  Transactions on Information Theory}, vol.~57, no.~4, pp. 2400--2419, 2011.

\bibitem{NAD11}
B.~Nadler, ``On the distribution of the ratio of the largest eigenvalue to the
  trace of a wishart matrix,'' \emph{Journal of Multivariate Analysis}, vol.
  102, no.~2, pp. 363--371, 2011.

\bibitem{CAR08}
L.~S. Cardoso, M.~Debbah, P.~Bianchi, and J.~Najim, ``{Cooperative spectrum
  sensing using random matrix theory},'' in \emph{{IEEE} Pervasive Computing
  (ISWPC'08)}, Santorini, Greece, May 2008, pp. 334--338.

\bibitem{COU09b}
R.~Couillet and M.~Debbah, ``{A Bayesian framework for collaborative
  multi-source signal detection},'' \emph{{IEEE} Transactions on Signal
  Processing}, vol.~58, no.~10, pp. 5186--5195, Oct. 2010.

\end{thebibliography}

\begin{IEEEbiography}[{\includegraphics[width=1in,height=1.25in]{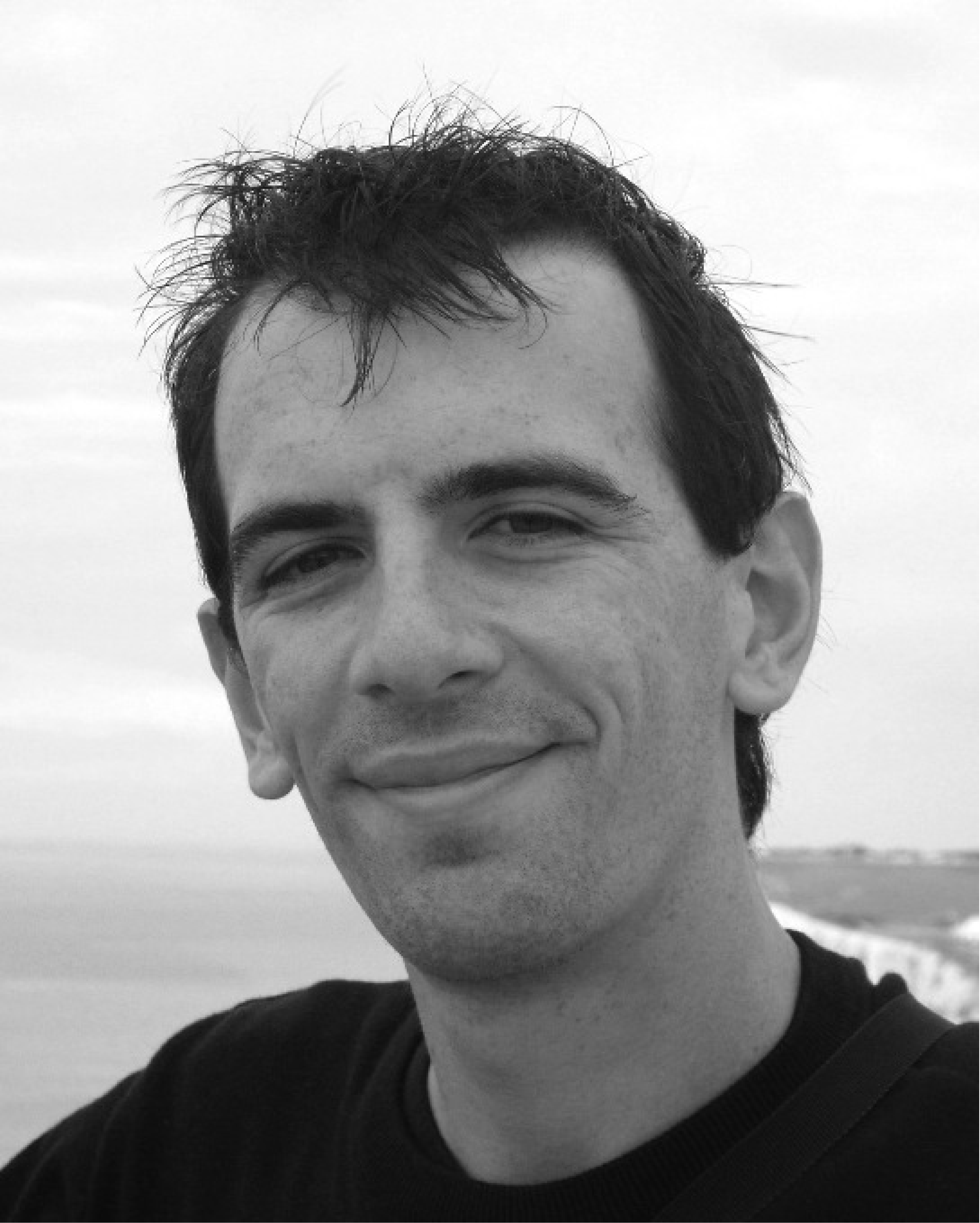}}]{Romain Couillet}
	received his MSc in Mobile Communications at the Eurecom Institute and his MSc in Communication Systems in Telecom ParisTech, France in 2007. From 2007 to 2010, he worked with ST-Ericsson as an Algorithm Development Engineer on the Long Term Evolution Advanced project, where he prepared his PhD with Sup\'elec, France, which he graduated in November 2010. He is currently an assistant professor in the Telecommunication department of Sup\'elec. His research topics are in information theory, signal processing, and random matrix theory. He is the recipient of the Valuetools 2008 best student paper award and of the 2011 EEA/GdR ISIS/GRETSI best PhD thesis award.
\end{IEEEbiography}

\begin{IEEEbiography}[{\includegraphics[width=1in,height=1.25in]{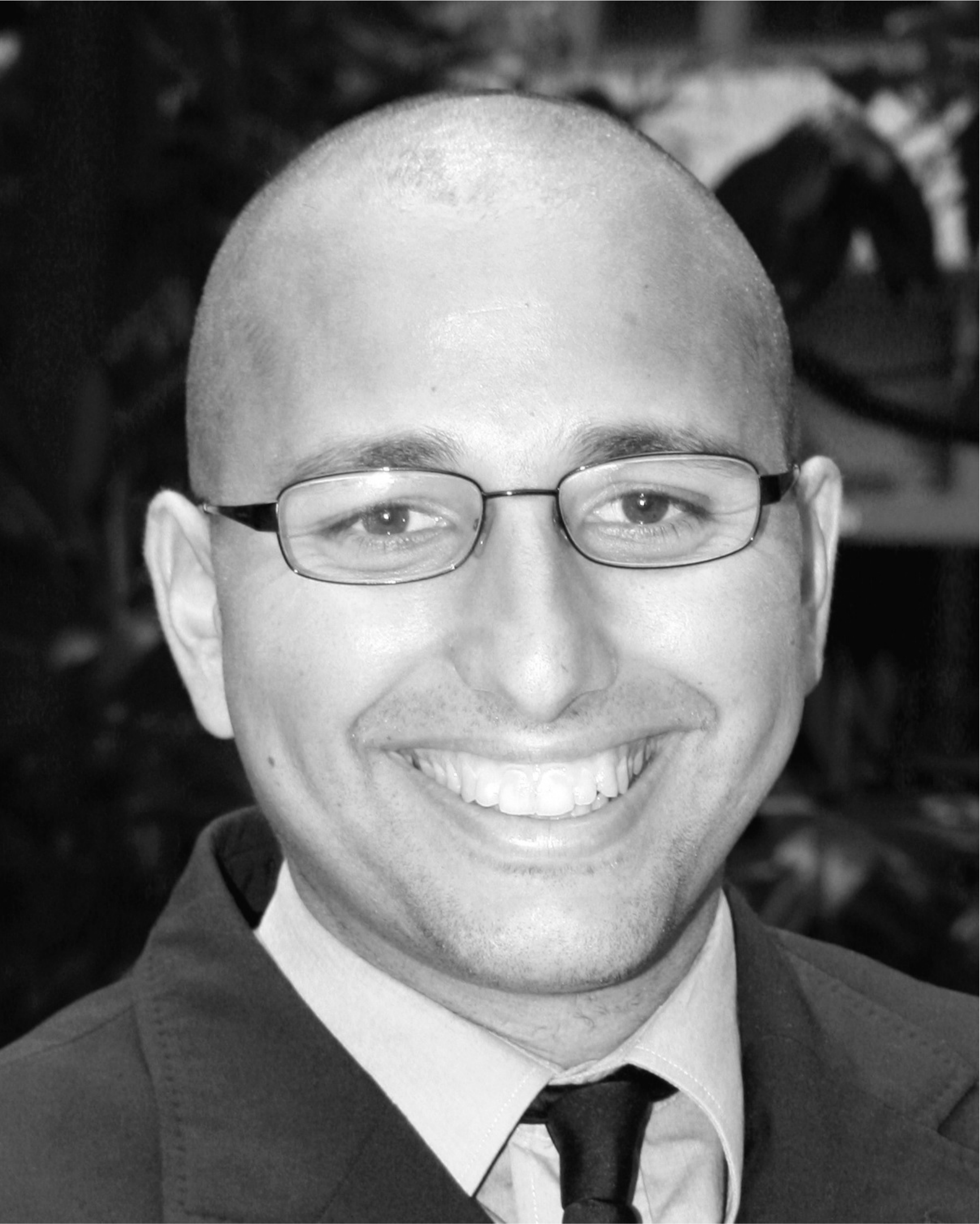}}]{M\'erouane Debbah}
	was born in Madrid, Spain. He entered the Ecole Normale Sup\'erieure de Cachan (France) in 1996 where he received his M.Sc and Ph.D. degrees respectively in 1999 and 2002. From 1999 to 2002, he worked for Motorola Labs on Wireless Local Area Networks and prospective fourth generation systems. From 2002 until 2003, he was appointed Senior Researcher at the Vienna Research Center for Telecommunications (FTW) (Vienna, Austria). From 2003 until 2007, he joined the Mobile Communications de-partment of the Institut Eurecom (Sophia Antipolis, France) as an Assistant Professor. He is presently a Professor at Sup\'elec (Gif-sur-Yvette, France), holder of the Alcatel-Lucent Chair on Flexible Radio. His research interests are in information theory, signal processing and wireless communications. Mérouane Debbah is the recipient of the ``Mario Boella'' prize award in 2005, the 2007 General Symposium IEEE GLOBECOM best paper award, the Wi-Opt 2009 best paper award, the2010  Newcom++ best paper award  as well as the Valuetools 2007, Valuetools 2008 and CrownCom2009 best student paper awards. He is a WWRF fellow.
\end{IEEEbiography}

\end{document}